\documentclass[12pt]{article}

\newcommand{\be}{\begin{equation}}
\newcommand{\ee}{\end{equation}}
\newcommand{\bi}[1]{\vspace{-3mm} \bibitem{#1}}

\usepackage{epsfig,amsmath,amssymb,graphics,graphicx} 

\begin{document}
\begin{center}

{\bf \large Non-Standard Extensions of Gradient Elasticity: \\
\vskip 3mm
Fractional Non-Locality, Memory and Fractality} \\

\vskip 7mm
{\bf \large Vasily E. Tarasov} \\
\vskip 3mm

{\it Skobeltsyn Institute of Nuclear Physics,\\ 
Lomonosov Moscow State University, \\
Moscow 119991, Russia} \\
{E-mail: tarasov@theory.sinp.msu.ru} \\

\vskip 7mm
{\bf \large Elias C. Aifantis}\footnote{Emeritus Professor of Engineering,
Michigan Tech, Houghton, MI 49931, USA\\
Distinguished Adjunct Professor of King Abdulaziz University,
Jeddah, 21589, SA} \\
\vskip 3mm

{\it Laboratory of Mechanics and Materials, \\ 
 Aristotle University of Thessaloniki, \\
Thessaloniki 54006, Greece} \\
{E-mail: mom@mom.gen.auth.gr} \\

\vskip 7mm
\begin{abstract}
Derivatives and integrals of non-integer order may 
have a wide application in describing complex properties of materials including long-term memory, 
non-locality of power-law type and fractality.
In this paper we consider extensions of elasticity theory 
that allow us to describe elasticity of 
materials with fractional non-locality, memory and fractality. 
The basis of our consideration is an extension of 
the usual variational principle for fractional non-locality and fractality. 
For materials with power-law non-locality described by 
Riesz derivatives of non-integer order, 
we suggest a fractional variational equation. 
Equations for fractal materials are derived by 
a generalization of the variational principle for fractal media.
We demonstrate the suggested approaches 
to derive corresponding generalizations 
of the Euler-Bernoulli beam and the Timoshenko beam equations 
for the considered fractional non-local and fractal models. 
Various equations for materials  with fractional non-locality, 
fractality and fractional acceleration are considered.
\end{abstract}

\end{center}



\section{Introduction}

Derivatives and integrals of non-integer orders 
\cite{SKM}-\cite{FC2} 
have wide applications in mechanics and physics  
\cite{CM}-\cite{IJMPB2013}.
The tools of fractional derivatives and integrals allow us 
to investigate the behavior of materials and systems 
that are characterized by power-law non-locality, 
power-law long-term memory and fractal properties.
As concluded from the above listed works,
there are different definitions of fractional derivatives
such as Riemann-Liouville, Riesz, Caputo, Gr\"unwald-Letnikov, 
Marchaud, Weyl, Sonin-Letnikov and others.
The specific choice of fractional derivatives for
a particular application, it thus depends on the taste
of the investigator and the nature of the material or 
system at hand.
Many properties of standard differentiation and integration
do not extend in the fractional case
and fractional counterpart of popular models
need to be rederived, each on individual basis.


Usually non-local continuum mechanics are treated 
with two approaches \cite{AA2011}:
The gradient elasticity theory (weak non-locality) and 
the integral non-local theory (strong non-locality).
The fractional calculus can, in fact, 
be used to formulate 
a generalization of non-local theory of elasticity in both forms:
fractional gradient elasticity  
(weak power-law non-locality) and 
fractional integral elasticity 
(strong power-law non-locality). 
In this paper, we consider fractional generalizations 
of the gradient elasticity theory only.
In particular, we suggest fractional generalizations
of a rather popular robust GRADELA model
proposed by Aifantis and co-workers \cite{A1992}-\cite{AMA2008}
for the following cases:

(1) The elasticity of materials with power-law non-locality 
that can be described by derivatives of non-integer order. 
Both 1D and 3D models are discussed.

(2) The elasticity of materials with power-law memory 
that can be described by fractional time derivatives 
for the internal inertia or
combined strain-acceleration fractional gradient terms.

(3) The elasticity of materials with fractal structure that
can be described by fractional integrals in the framework 
of fractional continuum models.

The basis of our consideration is an extension of 
the usual variational principle for materials 
with fractional non-locality, memory and fractality. 
For 3D spatial fractional models we also use the 
apparatus of fractional vector calculus.
An extension of the traditional calculus of variations 
for systems described by fractional derivatives 
was first proposed by Agrawal in \cite{Om-1} 
for the Riemann-Liouville derivatives.
Then it has been extended for other type of 
fractional derivatives \cite{RL-1}-\cite{Om-3},
and fractional integrals \cite{Int-1}.
For materials with power-law non-locality and memory, 
we suggest a new fractional variational principle 
for Lagrangians with Riesz fractional derivatives. 
A possible generalization of gradient elasticity theory 
for fractal materials  was alluded in \cite{ECA-2011}.
In this paper we describe fractal materials by using 
the fractional continuum formalism suggested in 
\cite{PLA2005-1,AP2005-2} 
(see also \cite{IJMPB2005-2}-\cite{MOS-1}).
To obtain governing equations for fractional integral
continuum models  of fractal materials, 
we employ a generalization of the holonomic variational principle 
suggested in \cite{Physica2005,MPLB2005-1}.
In this connection, we note that extremum and 
variational principles for non-gradient 
but fractal elastic materials 
within a fractional continuum model framework have been 
considered in \cite{MOS-1,MOS-2}.

The Euler-Bernoulli beam theory may be viewed as 
a benchmark example 
of the classical linear theory of elasticity. 
It provides tools for calculating the load-carrying and deflection characteristics of beams subjected to lateral loads only. 
In order to illustrate the implications of the suggested 
fractional approaches in this paper, 
we use a variational principle to derive the corresponding  generalizations of the static and dynamic Euler-Bernoulli 
beam model, as well as that of the Timoshenko beam model 
for the fractional non-local and fractal cases. 
Solutions to some of these equations for fractional non-local 
and fractal beams are considered.



Next, we list some non-standard generalizations of 
constitutive relations for gradient elasticity models.
First, we recall the linear elastic constitutive relations
for isotopic and homogeneous bodies, i.e.,
\be \label{H-0}
\sigma_{ij} = \lambda \varepsilon_{kk} \delta_{ij} + 2 \mu \varepsilon_{ij} ,
\ee
where $\sigma_{ij}$ is the stress tensor, $\varepsilon_{ij}$
is the strain tensor, whereas
$\lambda$ and $\mu$ are the Lame coefficients.
In \cite{A1992}-\cite{RA1993} it was suggested a generalization
of the constitutive relations (\ref{H-0}) by 
a gradient modification that contains the Laplacian $\Delta$ 
in the form
\be \label{H-1}
\sigma_{ij} = \Bigl( \lambda  \varepsilon_{kk} \delta_{ij} + 2 \mu \varepsilon_{ij} \Bigr)
- l^2_s \, \Delta \, \Bigl( \lambda  \varepsilon_{kk} \delta_{ij} + 2 \mu \varepsilon_{ij} \Bigr) ,
\ee
where $l_s$ is an internal length scale parameter \cite{AA2011}.
To describe complex materials characterized by 
non-locality of power-law type, long-term memory,  and fractality, we should further generalize Eq. (\ref{H-0}) 
and its gradient counterpart given by Eq. (\ref{H-1}).
In this paper, we consider the following non-standard generalizations of the gradient stress-strain relation.

(1) The fractional gradient elasticity models 
with power-law non-locality  
\be \label{Gen-1a}
\sigma_{ij} = \Bigl( \lambda  \varepsilon_{kk} \delta_{ij} + 2 \mu \varepsilon_{ij} \Bigr)
- l^2_s (\alpha) \, (- \, ^R\Delta)^{\alpha/2} \, \Bigl( \lambda  \varepsilon_{kk} \delta_{ij} + 2 \mu \varepsilon_{ij} \Bigr) ,
\ee
where $(- \, ^R\Delta)^{\alpha/2}$ is the fractional generalization of the Laplacian in the Riesz form, and
\be \label{Gen-1b}
\sigma_{ij} = \Bigl( \lambda  \varepsilon_{kk} \delta_{ij} + 2 \mu \varepsilon_{ij} \Bigr)
- l^2_s (\alpha) \, ^C\Delta^{\alpha}_W \, \Bigl( \lambda  \varepsilon_{kk} \delta_{ij} + 2 \mu \varepsilon_{ij} \Bigr) ,
\ee
where $\, ^C\Delta^{\alpha}_W$ is the fractional Laplacian in the Caputo form.

(2) The combined fractional strain gradient-internal 
inertia model with power-law memory and non-locality
\be \label{Gen-2a}
\sigma_{ij} = \Bigl( \lambda  \varepsilon_{kk} \delta_{ij} + 2 \mu \varepsilon_{ij} \Bigr)- \Bigl( l^2_s (\alpha) \, (- \, ^R\Delta)^{\alpha/2} \, + l^2_d (\beta) \, (\, ^RD^{\beta}_t)^2 \Bigr) \, \Bigl( \lambda  \varepsilon_{kk} \delta_{ij} + 2 \mu \varepsilon_{ij} \Bigr) ,
\ee
where $(\, ^RD^{\beta}_t)^2$ is the square of the derivative 
of non-integer order $\beta$ with respect to time $t$, which
describes acceleration with power-law memory. 

(3) The gradient elasticity models for fractal materials 
\be \label{Gen-3}
\sigma_{ij} = \Bigl( \lambda  \varepsilon_{kk} \delta_{ij} + 2 \mu \varepsilon_{ij} \Bigr)
- l^2_F(D,d) \, \Delta^{(D,d)} \, \Bigl( \lambda  \varepsilon_{kk} \delta_{ij} + 2 \mu \varepsilon_{ij} \Bigr) ,
\ee
where $\Delta^{(D,d)}$ is the "fractal-Laplacian"
that takes into account the power-law density of states 
of the fractal medium under consideration.

The plan of the paper is as follows:
In Section 2, we consider one-dimensional (1D) 
fractional gradient elasticity models. 
A variational principle for these models is suggested. 
Fractional Euler-Bernoulli and Timoshenko beam equations are derived.
Solutions for the fractional static and dynamic Euler-Bernoulli 
beam governing  equations are proposed. 
Corresponding fractional beam models with combined 
strain-internal inertia gradient terms are also considered.
Moreover, solutions of the relevant generalized equation and 
dispersion law for this model are derived.
In Section 3, three-dimensional (3D) fractional gradient elasticity models are formulated and discussed. 
In particular, 3D problems with spherical symmetry based 
on the Riesz fractional derivative are considered. 
In addition, fractional 3D gradient elasticity models based on 
fractional vector calculus are suggested. 
The operator split method for solving the relevant 
fractional gradient elasticity equations is formulated. 
To illustrate the potential of this method,
a simple fractional gradient model is considered as 
an application and an explicit solution is provided.
In Section 4, some basic concepts for extending
gradient elasticity models to fractal media are suggested. 
The equilibrium equations for fractal materials 
are first derived.
A variational principle for obtaining gradient elasticity 
equations for fractal materials is then proposed.
Finally, in Section 5, generalizations of the Euler-Bernoulli 
and Timoshenko beam equations for fractal materials and 
the corresponding equations for the combined strain-acceleration gradients fractal beam models are derived.


\newpage
\section{Fractional 1D gradient elasticity}

Fractional elasticity models are those for which   
non-locality of power-law type is described by using 
derivatives and integrals of non-integer order.
We can derive such phenomenological fractional elasticity models 
by using a variational principle for a Lagrangian with
fractional derivatives.
A generalization of the traditional calculus of variations 
for systems described by Riemann-Liouville fractional derivatives
has been suggested by Agrawal in \cite{Om-1}.
Then, extensions of variational calculus for 
the Riemann-Liouville derivatives \cite{RL-1}, 
the Caputo derivative \cite{Cap-1}-\cite{Cap-3},
the Hadamard derivative \cite{Om-2}, 
the Riesz derivative \cite{Om-3}, as well as
fractional integrals \cite{Int-1}, have been derived.

If we use the fractional derivatives of 
Riemann-Liouville, Caputo, Liouville, Marchaud,
then we should take into account 
the left-sided and the right-sided fractional derivatives
in the Lagrangian.
The correspondent fractional Euler-Lagrange equations contain 
the left-sided and the right-sided fractional derivatives also.
In addition, the integration by parts, which is used 
in the derivation of the Euler-Lagrange equations from 
the variational principle, transforms the left-sided derivatives 
into the right-sided (see Eq. 2.64 of \cite{SKM}).
As a result, we obtain a mixture of left-sided and 
the right-sided derivatives in the equations of motion.
Unfortunately, these Euler-Lagrange equations 
can be solved for a very narrow class of Lagrangians only.

In this paper, we suggest a fractional variational principle 
for systems that are described by 
Riesz fractional derivatives \cite{SKM,KST}.   
The suggested principle differs from 
the one proposed in \cite{Om-3}.
We take advantage of the fact that the Riesz derivative 
does not involve two forms, i.e., 
left-sided and right-sided derivatives.
In addition, integration by parts transforms 
the Riesz fractional derivative into itself. 
The corresponding fractional Euler-Lagrange equations 
can be solved for a wide class of Lagrangians that describe
nonlocal materials by the methods described in \cite{KST}.
Moreover, the Riesz fractional derivatives 
naturally arise in the elasticity theory  based on 
lattice models \cite{CEJP2013}-\cite{MOM2014}.
As an example, we derive the fractional gradient generalization of the Euler-Bernoulli beam model and provide
some general solutions of the corresponding equations 
for both static and dynamics configurations.

\subsection{Fractional 1D gradient elasticity from variational principle}

To generalize standard variational principles for fractional
nonlocal models, we write all expressions
in dimensionless coordinate variables.
We can introduce the dimensionless variables 
$x_k = x^{\prime}_k / l_0 , \quad {\bf r}={\bf r}^{\prime}/l_0$, 
where $l_0$ is a characteristic scale.
This allows us to have usual physical dimensions of 
measured quantities. 

The equation for the fractional gradient elasticity can be derived as the Euler-Lagrange equation of the following action
\be \label{action-1}
S [w] = \int dt \int dx \ 
\mathcal{L} (w, D^1_t w, \, ^RD^{\alpha_1}_x w , \, ^RD^{\alpha_2}_x w) ,
\ee
where $\mathcal{L}(w, D^1_t w, 
\, ^RD^{\alpha_1}_x w , \, ^RD^{\alpha_2}_x w)$ is the Lagrangian defining the 1D fractional elasticity model, $w=w(x,t)$ denotes 
the displacement field, and $x$ is the dimensionless coordinate.

The variation of the action functional (\ref{action-1}) with respect to $w(x,t)$ 
and its derivatives is given by
\[ \delta S [w] = 
\int dt \int dx \,  \delta \mathcal{L} = \int dt \, \int dx \, 
\Bigl[ \frac{\partial\mathcal{L}}{\partial w} \delta w + 
 \left(\frac{\partial \mathcal{L}}{\partial D^1_t w}\right) \, \delta (D^1_t w) + \]
\be \label{VarS-1}
+ \left(\frac{\partial \mathcal{L}}{\partial \, ^RD^{\alpha_1}_x w}\right) \, \delta (\, ^RD^{\alpha_1}_x w)  
+  \left(\frac{\partial \mathcal{L}}{\partial \, ^RD^{\alpha_2}_x w}\right) \, \delta (\, ^RD^{\alpha_2}_x w)  \Bigr] ,
\ee
where, in the absence of non-holonomic constraints, 
the variation and fractional derivatives commute, i.e. 
\[
\delta (D^1_t w) = D^1_t (\delta w) , \quad
\delta (\, ^RD^{\alpha_1}_x w)  = \, ^RD^{\alpha_1}_x (\delta w) , \quad
\delta (\, ^RD^{\alpha_2}_x w)  = \, ^RD^{\alpha_2}_x (\delta w)  .
\]


In order to utilize the fractional variational principle,
we should perform the operation of integration by parts.
Unfortunately, integration by parts transforms 
left-sided derivatives into right-sided
ones for the most commonly used types of fractional derivatives.
For the Liouville fractional derivatives
\be \label{Liouv} 
(\, ^L{\bf D}^{\alpha}_{\pm} f)(x)
= \frac{(-1)^n}{\Gamma(n-\alpha)} \frac{d^n}{dx^n}
\int^{\infty}_0 \frac{f(x \mp z)}{z^{\alpha+1-n}} dz  ,
\ee 
the integration by parts 
(see Eq. 5.17 in Section 5.1 of \cite{SKM}) has the form
\be
\int^{+\infty}_{-\infty}  f(x) \, (\, ^LD^{\alpha}_{+} g)(x) \, dx =
\int^{+\infty}_{-\infty}  (\, ^LD^{\alpha}_{-} f)(x) \, g(x) \, dx .
\ee
For the Marchaud fractional derivatives, which is defined by
\be \label{Mar} 
(\, ^M{\bf D}^{\alpha}_{\pm} f)(x)
= \frac{\alpha}{\Gamma(1-\alpha)}
\int^{\infty}_0 \frac{f(x)-f(x \mp z)}{z^{\alpha+1}} dz  ,
\ee
the integration by parts (see Eq. 6.27 in Corollary 2 
of Theorem 6.2 of \cite{SKM}) has the form
\be \label{MarD}
\int^{+\infty}_{-\infty} f(x) \, (\, ^MD^{\alpha}_{+} g)(x) \, dx =
\int^{+\infty}_{-\infty} (\, ^MD^{\alpha}_{-} f)(x) \, g(x) \, dx .
\ee
This relation is valid for functions 
$f(x) \in L_s (\mathbb{R})$,
$f(x) \in L_t (\mathbb{R})$, such that
$(\, ^MD^{\alpha}_{+} g)(x) \in L_p(\mathbb{R})$ and
$(\, ^MD^{\alpha}_{-} f)(x) \in L_r(\mathbb{R})$,
where $p>1$, $r>1$, 
\[ \frac{1}{p}+ \frac{1}{r}=1+\alpha, \quad 
\frac{1}{s}=\frac{1}{p} - \alpha, \quad 
\frac{1}{t}=\frac{1}{r} - \alpha . \]

We suggest the use of Riesz fractional derivatives.
It is known (see Section 20.1 of \cite{SKM}) that
the connection of the Riesz fractional derivative
to the Marchaud fractional derivatives has the form
\be \label{dif-GLR}
(\, ^RD^{\alpha}_x f)(x) = \frac{1}{2 \cos (\alpha \pi/2)} \, 
\Bigl(  (\, ^MD^{\alpha}_{+} f)(x) + (\, ^MD^{\alpha}_{-} f)(x)\Bigr) ,
\ee
where $\alpha>0$, and $\alpha \ne 1,2,3,...$.
Here $^RD^{\alpha}_x$ is the Riesz fractional derivative
defined by the equation
\be \label{dif-GLR2} 
(\, ^RD^{\alpha}_x f)(x) = 
-\frac{\alpha}{2 \Gamma(1-\alpha) \cos (\alpha \pi/2)}
\int^{\infty}_0 \frac{f(x+z)-2 f(x)+f(x-z)}{z^{\alpha+1}} dz , 
\ee
where $x \in \mathbb{R}$.
Note that the Riesz derivative for an integer $\alpha=2n$ gives\be
(\, ^RD^{2n}_x f)(x) = \, (-1)^n \, D^{2}_x f(x) ,
\ee
where $n \in \mathbb{N}$, i.e.
\be \label{Integer-1}
^RD^{2}_x  = - \, D^{2}_x , \quad \, ^RD^{4}_x  = 
\, D^{4}_x , \quad \, ^RD^{6}_x  = - \, D^{6}_x . 
\ee
Using relations (\ref{dif-GLR}) and (\ref{MarD}), 
we obtain the equation of the integration by parts for 
the Riesz fractional derivative (\ref{dif-GLR})
in the form
\be \label{IPP}
\int^{+\infty}_{-\infty} f(x) \, (\, ^RD^{\alpha}_{x} g)(x) \, dx =
\int^{+\infty}_{-\infty} (\, ^RD^{\alpha}_{x} f)(x) \, g(x) \, dx .
\ee
As a result, integration by parts in Eq. (\ref{IPP}) 
does not change the type of derivative, and also 
does not change the sign in front of the integral.


Using the integration by parts given by Eq. (\ref{IPP}), 
we can rewrite the variation in Eq. (\ref{VarS-1}) as
\be \label{VarS-2}
\delta S [w] = \int dt \, \int dx \, 
\Bigl[ \frac{\partial\mathcal{L}}{\partial w} \delta w - 
D^1_t \, \left(\frac{\partial \mathcal{L}}{\partial D^1_t w}\right) + 
\, ^RD^{\alpha_1}_x  \, \left(\frac{\partial \mathcal{L}}{\partial \, ^RD^{\alpha_1}_x w}\right)   
+ \, ^RD^{\alpha_2}_x \, \left(\frac{\partial \mathcal{L}}{\partial \, ^RD^{\alpha_2}_x w}\right) \Bigr] \, \delta w  .
\ee
Then, the stationary action principle in the form of 
the holonomic variational equation 
\[ \delta S [w] =0 \]
yields the equation
\be \label{ELE-1}
\frac{\partial\mathcal{L}}{\partial w} - 
D^1_t \, \left(\frac{\partial \mathcal{L}}{\partial D^1_t w}\right)  
+ \, ^RD^{\alpha_1}_x \, \left(\frac{\partial \mathcal{L}}{\partial \, ^RD^{\alpha_1}_x w}\right) 
+ \, ^RD^{\alpha_2}_x \, \left(\frac{\partial \mathcal{L}}{\partial \, ^RD^{\alpha_2}_x w}\right) = 0 .
\ee
This is the fractional Euler-Lagrange equation for the model described by the Lagrangian $\mathcal{L} = \mathcal{L}(w, D^1_t w, 
\, ^RD^{\alpha_1}_x w , \, ^RD^{\alpha_2}_x w)$.
In the next section, we use this equation to establish
a fractional generalization of the Euler-Bernoulli beam model.

\subsection{Fractional Euler-Bernoulli beam equation from variational principle}

The Lagrangian of Euler-Bernoulli beams with gradient power-law non-locality has the form
\[ \mathcal{L}(w, D^1_t w, 
\, ^RD^{\alpha_1}_x w , \, ^RD^{\alpha_2}_x w) = \frac{1}{2} \mu \,
\left( D^1_t w(x,t) \right)^2 -
\frac{1}{2} (E\, I) \, \left( \, ^RD^{\alpha_1}_x w(x,t) \right)^2 -\]
\be \label{Lagr-Frac}
- \frac{1}{2} (E\, I) \, l^2_s (\alpha_2) 
\left( \, ^RD^{\alpha_2}_x w(x,t) \right)^2 + q(x,t) w(x,t) .
\ee
The curve $w(x)=u_y(x)$ describes the deflection of the beam in the $y$ direction at some position $x$. 
As we have already noted, $x$ and $l^2_s (\alpha_2)$ 
are dimensionless values. 
The first term represents the kinetic energy, 
where $\mu = \rho \, A$ is the mass per unit length; 
the second term describes the potential energy due 
to internal forces (when considered with a negative sign); and 
the third term is the potential energy due to the external load $q(x)$.
Note that in the Lagrangian of Eq. (\ref{Lagr-Frac}) 
the second term has a negative sign, since integration 
by parts in Eq. (\ref{IPP}) does not change the sign  
in front of the integral,
in contrast to the standard case.

For the usual case of $\alpha_1=2$ and $\alpha_3=3$, 
the Lagrangian given by Eq. (\ref{Lagr-Frac}) is
\[ \mathcal{L}(w, D^1_t w, D^{2}_x w , D^{3}_x w) = 
\frac{1}{2} \mu \, \left( D^1_t w(x,t) \right)^2 -
\frac{1}{2} (E\, I) \, \left( D^{2}_x w(x,t) \right)^2 + \]
\be \label{Lagr-Frac-Int}
+ \frac{1}{2} (E\, I) \, l^2_s \, 
\left( D^{3}_x w(x,t) \right)^2 + q(x,t) w(x,t) .
\ee

For the fractional case, the Lagrangian (\ref{Lagr-Frac}) 
leads to the expressions
\be \label{DL-1} 
\frac{\partial\mathcal{L}}{\partial w} = q(x,t)  \qquad
\frac{\partial \mathcal{L}}{\partial D^1_t w(x,t) } =  
\mu \, D^1_t w(x,t) , 
\ee
\be \label{DL-2} 
\frac{\partial \mathcal{L}}{\partial \, ^RD^{\alpha_1}_x w(x,t)} = - (E\, I) \, ^RD^{\alpha_1}_x w(x,t)  , \qquad
\frac{\partial \mathcal{L}}{\partial \, ^RD^{\alpha_2}_x w(x,t)} = - (E\, I) \, l^2_s(\alpha_2) \, ^RD^{\alpha_2}_x w(x,t).
\ee
Substitution of Eqs. (\ref{DL-1}) and (\ref{DL-2}) into the Euler-Lagrange equation (\ref{ELE-1}) gives
\be \label{FEB-1}
\mu \, D^2_t w +
\, ^RD^{\alpha_1}_x \, \Bigl( (E\, I) \, (\, ^RD^{\alpha_1}_x) w \Bigr) +
\, ^RD^{\alpha_2}_x \, \Bigl( (E\, I) \, l^2_s (\alpha_2) \,  ^RD^{\alpha_2}_x w \Bigr) 
- q(x,t) = 0 ,
\ee
which is the governing equation for the dynamics of a fractional non-local Euler-Bernoulli beam.

When the beam is homogeneous, $E$ and $I$ are independent of $x$, and the fractional Euler-Bernoulli beam equation assumes
the simpler form
\be \label{FEB-2}
\mu \, D^2_t w + 
(E\, I) \, (\, ^RD^{\alpha_1}_x)^2 w +
(E\, I) \, l^2_s (\alpha_2) \, (\, ^RD^{\alpha_2}_x)^2 w 
- q(x,t) = 0 .
\ee
For a wide class of functions $w(x)$
the properties of the fractional Riesz derivatives allows 
us to write Eq. (\ref{FEB-2}) as
\be \label{FEB-2b}
\mu \, D^2_t w + 
(E\, I) \, ^RD^{2\alpha_1}_x w +
(E\, I) \, l^2_s (\alpha_2) \, ^RD^{2\alpha_2}_x w 
- q(x,t) = 0 .
\ee
In general, we should consider an effective source term $q_{eff}(x)$  instead of $q(x)$, where $q_{eff}(x)$  
contains the function $q(x)$ and deviations from the semigroup property for the Riesz derivatives as 
described for the fractional gradient model 
with Caputo derivatives dealt with in \cite{TA2014}.

For materials without non-locality and memory, 
we have $\alpha_1=2$, $\alpha_2 =3$,
and then Eq. (\ref{FEB-2}) obtains the form
\be \label{EB-1}
\mu \, D^2_t w + E\, I \, D^4_x w - E\, I \, l^2_s \, D^6_x w 
- q(x,t) = 0 .
\ee
This is the gradient elasticity Euler-Bernoulli beam equation
derived earlier in \cite{AA2011} for the case of 
integer-order derivatives and non-fractal media.

\subsection{Solution of fractional static Euler-Bernoulli beam equation}

For the static case ($D^1_t w=0$ and $q(x,t)=q(x)$), equation (\ref{FEB-2b}) has the form
\be \label{FEB-2st2}
^RD^{2\alpha_1}_x w +
l^2_s (\alpha_2) \, ^RD^{2\alpha_2}_x w =  (E\, I)^{-1} \, q(x) .
\ee
Using Corollary 5.14 of \cite{KST}, we can state that 
a particular solution of equation (\ref{FEB-2st2}) is
\be
w(x)= (E\, I)^{-1} \, \int^{+\infty}_{-\infty} G_{2\alpha_1,2\alpha_2} (x- x^{\prime}) \, q(x^{\prime}) dx^{\prime} ,
\ee
where $G_{\alpha_1,\alpha_2}(x)$ is a Green's type function 
of the form
\be \label{GG2}
G_{2\alpha_1,2\alpha_2}(x) =  
\int^{\infty}_0 
\frac{ \cos ( \lambda |x| )}{\lambda^{2\alpha_1} + l^2_s (\alpha_2) \lambda^{2\alpha_2}}  \, d \lambda .
\ee 
Here $\alpha_1>0$, $\alpha_2>0$ and $ l^2_s (\alpha_2) \ne 0$.

For a point load of intensity $q_0$, i.e. 
a load $q(x)$ of the form  \cite{LL}
\be \label{deltaf}
q(x) = q_0 \, \delta (x) ,
\ee
where $\delta (x)$ denotes the Dirac delta-function, 
the displacement field $w (x)$  has a simple form 
$w (x) = (q_0/E\, I) \, G_{2\alpha_1,2\alpha_2}(x)$
given by expression
\be \label{phi-Gb}
w (x) = 
\frac{2 \, q_0}{\pi \, E\, I} \, \int^{\infty}_0 
\frac{ \cos ( \lambda |x| )}{\lambda^{2\alpha_1} + l^2_s (\alpha_2) \lambda^{2\alpha_2}}  \, d \lambda ,
\ee
where the definition given by Eq. (\ref{GG2})
for $G_{2\alpha_1,2\alpha_2}(x)$ has been used. 
For the usual non-fractional case, 
the solution of the static Euler-Bernoulli beam equation
with the external point-load is given by Eq. (\ref{phi-Gb}) 
with $\alpha_1=2$ and $\alpha_3=3$.

\subsection{Solution of fractional dynamic Euler-Bernoulli beam equation}

For a plane wave traveling in a fractional non-local material
with frequency $\omega$, the governing fractional equation is
\be \label{FEB-2d}
- \mu \, \omega^2 \, w_p(x) + 
(E\, I) \, (\, ^RD^{\alpha_1}_x)^2 w_p(x) +
(E\, I) \, l^2_s (\alpha_2) \, (\, ^RD^{\alpha_2}_x)^2 w_p(x) 
- q_p(x) = 0 ,
\ee
where $w(x,t) = e^{- i \omega \, t} \, w_p(x)$, and
we have also used the notation
$q(x,t) = e^{- i \omega \, t} \, q_p(x)$. 
For a wide class of functions $w_p(x)$,
Eq. (\ref{FEB-2d}) can be expressed as
\be \label{FEB-2d2}
^RD^{2\alpha_1}_x w_p(x) +
 l^2_s (\alpha_2) \, ^RD^{2\alpha_2}_x w_p(x) 
- \frac{\mu \, \omega^2}{E\, I} \, w_p(x) 
= (E\, I)^{-1} \, q_p(x) .
\ee

Using Theorem 5.24 of \cite{KST}, we can obtain 
a particular solution of Eq. (\ref{FEB-2d2}) as
\be \label{Solut-1}
w_p(x,\omega)= (E\, I)^{-1} \, \int^{+\infty}_{-\infty} G_{2\alpha_1,2\alpha_2} (x- x^{\prime}, \omega) q_p(x^{\prime}) dx^{\prime} ,
\ee
where $ G^{(\omega)}_{2\alpha_1,2\alpha_2}(x)$ 
is a Green's type function of the form
\be
G_{2\alpha_1,2\alpha_2} (x,\omega) = 2 \, 
\int^{\infty}_0 
\frac{ \cos ( \lambda \, |x|)}{\lambda^{2\alpha_1} + l^2_s (\alpha_2) \lambda^{2\alpha_2} - \mu \, \omega^2/(E\, I)}  \, d \lambda .
\ee 
Here $\alpha_1>0$, $\alpha_2>0$, $l^2_s (\alpha_2) \ne 0$ and $\mu \, \omega^2 \ne 0$, $\mu=\rho\, A$.
For the point-load case ({\ref{deltaf}}), 
the solution given by Eq. (\ref{Solut-1}) 
is reduced to
\be \label{Solut-2}
w_p(x,\omega)= \frac{2 q_0}{E\, I} \,
\int^{\infty}_0 
\frac{ \cos ( \lambda \, |x|)}{\lambda^{2\alpha_1} + l^2_s (\alpha_2) \lambda^{2\alpha_2} - \mu \, \omega^2/(E\, I)}  \, d \lambda .
\ee 
For the usual non-fractional case, 
the solution of dynamic Euler-Bernoulli beam equation
with an external point-load is given by Eq. (\ref{Solut-2}) 
with $\alpha_1=2$ and $\alpha_3=3$.

\subsection{Fractional gradient Timoshenko beam equations}

In the Timoshenko beam theory the displacement vector 
${\bf u}(x,y,z,t)$ of the beam is assumed to be given by
\be 
u_x(x,y,z,t) = -z \, \varphi(x,t) \, \quad 
u_y(x,y,z,t) = 0 , \quad 
u_z(x,y,t) = w(x,t) ,
\ee
where $(x,y,z)$ are the coordinates of a point in the beam, 
($u_x$, $u_y$, $u_z$) are the corresponding components 
of the displacement vector, 
$\varphi=\varphi(x,t)$ is the angle of rotation of the normal 
to the mid-surface of the beam, 
and $w=w(x,t)$ is the displacement of the mid-surface in the $z$-direction.

To obtain a fractional generalization of the relevant gradient 
beam equation we use a fractional variational principle and a generalization of the Timoshenko beam Lagrangian. 
The appropriate form of such Lagrangian  
with fractional gradient non-locality, is
\[ \mathcal{L} 
= \frac{1}{2} \rho \, I \, \left( D^1_t \varphi(x,t) \right)^2 +
\frac{1}{2} \rho \, A \, \left( D^1_t w(x,t) \right)^2 -
\]
\[ 
- \frac{1}{2} (k G A) \, \left( \, ^RD^{\alpha_1}_x w(x,t) - \varphi(x,t)  \right)^2 
 - \frac{1}{2} (E\, I) \, \left( \, ^RD^{\beta_1}_x \varphi(x,t) \right)^2 -
\]
\be \label{Lagr-FGTB}
- \frac{1}{2} (k G A) \, l^2_s \,
\left( \, ^RD^{\alpha_2}_x w(x,t) - \, ^RD^{\beta_1}_x \varphi) \right)^2  
- \frac{1}{2} (E\, I) \, l^2_s \,
\left( \, ^RD^{\beta_2}_x \varphi (x,t) \right)^2 ,
\ee
where $(x,y,z)$ are dimensionless coordinates.
Note again that we use dimensionless coordinates. such that
the relevant quantities of fractional models have 
the same physical dimension as 
as corresponding one for non-fractional models. 

Then, in view of the expressions
\be \label{FTB-1} 
\frac{\partial \mathcal{L}}{\partial D^1_t w } =  
\rho \, A \, D^1_t w ,  \quad
\frac{\partial \mathcal{L}}{\partial \, ^RD^{\alpha_1}_x w} = -\, k\, G\, A  \, \Bigl( \, ^RD^{\alpha_1}_x  w - \varphi \Bigr) ,
\ee
\be \label{FTB-2} 
\frac{\partial\mathcal{L}}{\partial \varphi } = k\, G\, A  \, 
\Bigl( \, ^RD^{\alpha_1}_x  w - \varphi \Bigr)  \qquad
\frac{\partial \mathcal{L}}{\partial D^1_t \varphi } =  
\rho \, I \, D^1_t \varphi , 
\ee
\be \label{FTB-3} 
\frac{\partial \mathcal{L}}{\partial \, ^RD^{\beta_1}_x \varphi } = - E\, I\, ^RD^{\beta_1}_x  \varphi + 
l^2_s \, k \, G \, A \, ^RD^{\alpha_2}_x  w 
- l^2_s \, k \, G \, A \, ^RD^{\beta_1}_x  \varphi ,
\ee
\be \label{FTB-4} 
\frac{\partial\mathcal{L}}{\partial \, ^RD^{\alpha_2}_x w } = 
- l^2_s\,  k\, G\, A  \, 
\Bigl( \, ^RD^{\alpha_2}_x  w - \, ^RD^{\beta_1}_x  \varphi \Bigr) , \qquad
\frac{\partial \mathcal{L}}{\partial \, ^RD^{\beta_2}_x \varphi } =  
- \, l^2_s\, E \, I \, ^RD^{\beta_2}_x \varphi ,
\ee
the stationary action principle gives the 
following Euler-Lagrange equations
\be \label{FTB-5}
\frac{\partial\mathcal{L}}{\partial w} - 
D^1_t \, \left(\frac{\partial \mathcal{L}}{\partial D^1_t w}\right)  
+\, ^RD^{\alpha_1}_x \, \left( \frac{\partial \mathcal{L}}{\partial \, ^RD^{\alpha_1}_x w}\right)  
+ \, ^RD^{\alpha_2}_x \, \left(\frac{\partial \mathcal{L}}{\partial \, ^RD^{\alpha_2}_x w}\right) 
= 0 ,
\ee
\be \label{FTB-6}
\frac{\partial\mathcal{L}}{\partial \varphi } - 
D^1_t \, \left(\frac{\partial \mathcal{L}}{\partial D^1_t \varphi }\right)  
+ \, ^RD^{\beta_1}_x \, \left(\frac{\partial \mathcal{L}}{\partial \, ^RD^{\beta_1}_x  \varphi }\right) 
+ \, ^RD^{\beta_2}_x \, \left(\frac{\partial \mathcal{L}}{\partial \, ^RD^{\beta_2}_x \varphi }\right) = 0 .
\ee
Equations (\ref{FTB-5}) and (\ref{FTB-6}) are the Euler-Lagrange equations for the fractional gradient elasticity model described by the Lagrangian (\ref{Lagr-FGTB}).

Substitution of Eqs. (\ref{FTB-1}) - (\ref{FTB-4}) into 
Eqs. (\ref{FTB-5}) - (\ref{FTB-6}) gives the following 
fractional gradient Timoshenko beam equations 
for the displacement $w=w(x)$ and 
the rotation $\varphi=\varphi(x)$,
\be \label{FGTBE-1}
\rho \, A \, D^2_t w = 
\, ^RD^{\alpha_1}_x \, \left( -\,  k\, G\, A  \, \Bigl( \, ^RD^{\alpha_1}_x  w - \varphi \Bigr) \right)  
+ \, ^RD^{\alpha_2}_x \, \left( - l^2_s\,  k\, G\, A  \, 
\Bigl( \, ^RD^{\alpha_2}_x  w - \, ^RD^{\beta_1}_x  \varphi \Bigr)\right) ,
\ee
\[
\rho \, I \, D^2_t \varphi = 
k\, G\, A  \, \Bigl( \, ^RD^{\alpha_1}_x  w - \varphi \Bigr)  
+ \, ^RD^{\beta_1}_x \, \left( - E\, I\, ^RD^{\beta_1}_x  \varphi + 
l^2_s \, k \, G \, A \, ^RD^{\alpha_2}_x  w 
- l^2_s \, k \, G \, A \, ^RD^{\beta_1}_x  \varphi \right) +
\]
\be \label{FGTBE-2}
+ \, ^RD^{\beta_2}_x \, \left(  - \, l^2_s\, E \, I \, ^RD^{\beta_2}_x \varphi  \right) .
\ee
For homogeneous materials, Eqs. (\ref{FGTBE-1}) 
and (\ref{FGTBE-2}) take the form
\be \label{FGTBE-3}
\rho \, A \, D^2_t w = -\,
k\, G\, A  \, ^RD^{\alpha_1}_x \Bigl( \, ^RD^{\alpha_1}_x  w - \varphi \Bigr)  
- l^2_s\,  k\, G\, A \, 
^RD^{\alpha_2}_x \Bigl( \, ^RD^{\alpha_2}_x  w - \, ^RD^{\beta_1}_x  \varphi \Bigr) ,
\ee
\[
\rho \, I \, D^2_t \varphi = 
k\, G\, A  \, \Bigl( \, ^RD^{\alpha_1}_x  w - \varphi \Bigr)  
- E\, I\, ^RD^{\beta_1}_x  \, ^RD^{\beta_1}_x  \varphi + 
l^2_s \, k \, G \, A \, ^RD^{\beta_1}_x \, ^RD^{\alpha_2}_x  w -
\]
\be \label{FGTBE-4}
- l^2_s \, k \, G \, A \, ^RD^{\beta_1}_x \, ^RD^{\beta_1}_x  \varphi 
- \, l^2_s\, E \, I \, ^RD^{\beta_2}_x  \, ^RD^{\beta_2}_x \varphi .
\ee

For a wide class of functions $w(x,t)$ and $\varphi(x,t)$, 
Eqs. (\ref{FGTBE-3}) and (\ref{FGTBE-4}) can be rewritten as
\be \label{FGTBE-5}
\rho \, A \, D^2_t w = -\,
k\, G\, A  \, \Bigl( \, ^RD^{2\alpha_1}_x  w - \, ^RD^{\alpha_1}_x \varphi \Bigr)  
- l^2_s\,  k\, G\, A  \, 
\Bigl( \, ^RD^{2 \alpha_2}_x  w - \, ^RD^{\alpha_2 + \beta_1}_x  \varphi \Bigr) ,
\ee
\[
\rho \, I \, D^2_t \varphi = 
k\, G\, A  \, \Bigl( \, ^RD^{\alpha_1}_x  w - \varphi \Bigr)  
- E\, I\, ^RD^{2\beta_1}_x  \varphi + 
l^2_s \, k \, G \, A \, ^RD^{\alpha_2 +\beta_1}_x  w -
\]
\be \label{FGTBE-6}
- l^2_s \, k \, G \, A \, ^RD^{2\beta_1}_x  \varphi 
- \, l^2_s\, E \, I \, ^RD^{2\beta_2}_x \varphi .
\ee
If $\alpha_1=\beta_1=1$, and $\alpha_2=\beta_2=0$, 
Eqs. (\ref{FGTBE-5})-(\ref{FGTBE-6}) reduce to 
the well-known Timoshenko beam equations.
If $\alpha_1=\beta_1=1$, and $\alpha_2=\beta_2=2$, 
Eqs. (\ref{FGTBE-5})-(\ref{FGTBE-6}) reduce to 
the form of the gradient generalization of 
the Timoshenko beam equations.
In general, the Riesz fractional derivatives do not commute and
\be
^RD^{\alpha}_x \, ^RD^{\beta}_x \ne \, ^RD^{\alpha+\beta}_x .
\ee
In this case, Eqs. (\ref{FGTBE-3})-(\ref{FGTBE-4})
give Eqs. (\ref{FGTBE-5})-(\ref{FGTBE-6})
with an additional term in the form of 
an effective source terms that   
contain the deviations from the semigroup property for the Riesz derivatives as it was described in \cite{TA2014}.

\subsection{Combined strain-acceleration fractional gradients beam model}

Let us now consider internal inertia effects,
i.e. effects of combined strain-acceleration gradients 
on fractional nonlocal beams.
We start with the governing equation of a gradient elasticity Euler-Bernoulli beam equation with 
internal inertia or acceleration gradients  \cite{AA2011}, i.e.,
\be \label{EB-2com1}
\rho \, A \, D^2_t w + E\, I \, D^4_x w 
- E\, I \, l^2_s \, D^6_x w 
+ \rho \, I \, l^2_d \, D^2_t \, D^4_x w 
- q(x,t) = 0 ,
\ee
where $(\rho, A, E, I)$ have their usual meaning,
($x$, $t$) are dimensionless variables, and
($l^2_s$, $l^2_d$) are scale parameters.
The fractional generalization of Eq. (\ref{EB-2com1}) 
can be written in the form
\be \label{FEB-2com2}
- \rho \, A \, ^RD^{2\beta}_t w +
E\, I \, ^RD^{2 \alpha_1}_x w +
E\, I \, l^2_s (\alpha_2) \, ^RD^{2 \alpha_2}_x w 
- \rho \, I  \, l^2_d(\alpha_3) \, ^RD^{2\beta}_t \, ^RD^{2\alpha_3}_x w
- q(x,t) = 0 ,
\ee
where $\, ^RD^{2\beta}_t$ is the Riesz fractional derivative \cite{KST} with respect to time.
Using Eq. (\ref{Integer-1}), Eq. (\ref{FEB-2com2}) 
with $\alpha_1=\alpha_3=2$, $\beta=1$ and $\alpha_2=3$ 
gives Eq. (\ref{EB-2com1}).

Equation (\ref{FEB-2com2}) can be obtained from the stationary 
action principle and the correspondent
fractional Euler-Lagrange equation
\be 
\frac{\partial\mathcal{L}}{\partial w} 
+ \, ^RD^{\beta}_t \, \left(\frac{\partial \mathcal{L}}{\partial \, ^RD^{\beta}_t w}\right) 
+ \, ^RD^{\alpha_3}_x \, ^RD^{\beta}_t \, \left(\frac{\partial \mathcal{L}}{\partial \, ^RD^{\alpha_3}_x \, ^RD^{\beta}_t w}\right) 
+ \, ^RD^{\alpha_1}_x \, \left(\frac{\partial \mathcal{L}}{\partial \, ^RD^{\alpha_1}_x w}\right) 
+ \, ^RD^{\alpha_2}_x \, \left(\frac{\partial \mathcal{L}}{\partial \, ^RD^{\alpha_2}_x w}\right) = 0 ,
\ee
where the Lagrangian 
\[ \mathcal{L}
=  \frac{1}{2} \, \rho \, A \,
\left( \, ^RD^{\beta}_t w(x,t) \right)^2 -
\frac{1}{2} \, E \, I  \, \left( \, ^RD^{\alpha_1}_x w(x,t) \right)^2 -\]
\be \label{Lagr-Frac2}
- \frac{1}{2} \, E \, I \, l^2 (\alpha_2) 
\left( \, ^RD^{\alpha_2}_x w(x,t) \right)^2  
+ \frac{1}{2} \rho \, I \, l^2_d (\alpha_3) 
\left( \, ^RD^{\alpha_3}_x \, ^RD^{\beta}_t  w(x,t) \right)^2  
+ q(x,t) w(x,t) ,
\ee
is used and Eq. (\ref{Integer-1}) is also taken into account.


In the above we use
the Riesz fractional derivatives (\ref{dif-GLR}) 
with respect to time to derive Eq.(\ref{FEB-2com2}) 
from a variational principle, instead of the Caputo derivatives 
that are commonly used.
Moreover, the Riesz fractional derivatives allow us to obtain 
a general harmonic solution of the combined strain-acceleration fractional gradient beam model, as we will see in the sequel.
At the same time, an interpretation of Riesz fractional derivatives with respect to time can be more complicated 
in comparison with the left-sided Caputo derivative.
In any case, the Riesz fractional time derivative describes 
a special form of power-law material memory (acceleration with memory) and deserves to be explored in its own right.

\subsection{Solution for the combined strain-acceleration fractional gradients beam model} 

Let us consider the Fourier transform ${\cal F}$ 
of the displacement field by utilizing the properties of     
the Riesz fractional derivative (see Property 2.34 in \cite{KST}) 
with respect to time
\be \label{Four}
\Bigl( {\cal F } \, ^RD^{\alpha}_t w (x,t) \Bigr)(k) = 
|k|^{\alpha} \, ( {\cal F } w (x,t) ) (x,\omega) ,
\ee
where $w(x,t)$ belongs to the space 
$C^{\infty}_0(\mathbb{R}^2)$ 
of infinitely differentiable functions on 
$\mathbb{R}^2$ with a compact support.
Then Eq. (\ref{FEB-2com2}) takes the form
\be \label{FEB-2com2-Omega1}
- \rho \, A \, |\omega|^{2\beta} \hat w 
+ E\, I \, ^RD^{2 \alpha_1}_x \hat w 
+ E\, I \, l^2_s (\alpha_2) \, ^RD^{2 \alpha_2}_x \hat w 
- \rho \, I \, l^2_d(\alpha_3) \, \omega^{2\beta} \, ^RD^{2\alpha_3}_x \hat w 
- \hat q(x,\omega) = 0 ,
\ee
where $\hat w (x,\omega) = ({\cal F } w (x,t)) (x,\omega)$
and  $\hat q (x,\omega) = ({\cal F } q (x,t)) (x,\omega)$.
By rewriting this equation in the form
\be \label{FEB-2com2-Omega2}
E\, I \, l^2_s (\alpha_2) \, ^RD^{2 \alpha_2}_x \hat w 
+ E\, I \, ^RD^{2 \alpha_1}_x \hat w 
- \rho \, I \, l^2_d(\alpha_3) \, \omega^{2\beta} \, ^RD^{2\alpha_3}_x \hat w 
- \rho \, A \, \omega^{2\beta} \hat w 
= \hat q(x,\omega) ,
\ee
we can solve it by using Theorem 5.24 of \cite{KST} 
with the coefficients
\be \label{all-a}
a_0 = \rho \, A \, \omega^{2\beta}, \quad
a_1 = \rho \, I \, l^2_d(\alpha_3) \, \omega^{2\beta} , \quad
a_2 = E\, I ,\quad
a_3 = E\, I \, l^2_s (\alpha_2) .
\quad
\ee


Noting that the Fourier transform of 
the Riesz fractional derivative with respect to coordinates 
is defined by
\be \label{FFL}
({\cal F} (\, ^RD^{2\alpha} \hat w(x,\omega))(k,\omega)= 
|k|^{2\alpha} \, ({\cal F}\hat w)(k,\omega) ,
\ee
and applying ${\cal F}$ to both sides 
of Eq. (\ref{FEB-2com2-Omega2})
by also using Eq. (\ref{FFL}), we obtain
\be
({\cal F} \hat w)(k,\omega) = \left( 
a_3 \, |k|^{2 \alpha_2} + a_2 \, |k|^{2 \alpha_1} - a_1 \, |k|^{2 \alpha_3} - a_0 \right)^{-1} 
({\cal F} \hat w)(k,\omega) .
\ee
Next, we define the fractional analogue of 
Green's function \cite{KST} as
\[
G_{\alpha}(x)= {\cal F}^{-1} \Bigl[ \left( a_3 \, |k|^{2 \alpha_2} + a_2 \, |k|^{2 \alpha_1} - a_1 \, |k|^{2 \alpha_3} - a_0 \right)^{-1} \Bigr] (x)=
\]
\be \label{FGF}
=\int_{\mathbb{R}} \left( a_3 \, |k|^{2 \alpha_2} + a_2 \, |k|^{2 \alpha_1} - a_1 \, |k|^{2 \alpha_3} - a_0 \right)^{-1} \
e^{ + i k x } \, d k ,
\ee
where $\alpha=(\alpha_1,\alpha_2,\alpha_3)$ is a multi-index.
Then, the following relation holds
\be \label{3-1}
\int_{\mathbb{R}^n} e^{  i ({\bf k},{\bf r}) } \, f(|{\bf k}|) \, d^n {\bf k}= 
\frac{(2 \pi)^{n/2}}{ |{\bf r}|^{(n-2)/2}} 
\int^{\infty}_0 f( \lambda) \, \lambda^{n/2} \, J_{n/2-1}(\lambda |{\bf r}|) \, d \lambda
\ee
for any function $f$
such that the integral in the right-hand side of Eq. (\ref{3-1}) is convergent (see Lemma 25.1 of \cite{SKM}).
Here $J_{\nu}$ is the Bessel function of the first kind and
we can use for $n=1$ the expression
\be
J_{-1/2} (z) = \sqrt{\frac{2}{\pi \, z}} \cos (z) .
\ee

Using Eq. (\ref{3-1}), the Green's function 
given by Eq. (\ref{FGF}) can be rewritten 
(see Theorem 5.22 of \cite{KST}) in the form 
\be \label{G-1}
G_{\alpha} ({\bf r}) = (2 \pi)^{1/2} |x|^{1/2} 
\int^{\infty}_0 \frac{\lambda^{1/2} \, J_{-1/2} (\lambda |x|) \, d \lambda}{a_3 \, \lambda^{2 \alpha_2} + a_2 \, \lambda^{2 \alpha_1} - a_1 \, \lambda^{2 \alpha_3} - a_0}  ,
\ee
where we used $n=1$.
If $\alpha_2 > 1$, and $a_k \ne 0$, 
then Eq. (\ref{FEB-2com2-Omega2}) is solvable \cite{KST}.
A particular solution of Eq. (\ref{FEB-2com2-Omega2}) can be represented as the
convolution of the functions $G(x)$ and $q(x)$, i.e.,
\be \label{w-G}
\hat w(x,\omega)= \int_{\mathbb{R}} G_{\alpha} (x - x^{\prime}) \, 
q (x,\omega) \, d x^{\prime} ,
\ee
where the Green's function $G_{\alpha}(z)$ 
is defined by Eq. (\ref{G-1}).

For the case $q (x,\omega)=q_0\delta(x)$, equation (\ref{w-G}) gives
\be \label{w-G2}
\hat w(x,\omega)= 2 q_0
\int^{\infty}_0 \frac{ \cos( \lambda |x|) \, d \lambda}{a_3 \, \lambda^{2 \alpha_2} + a_2 \, \lambda^{2 \alpha_1} - a_1 \, \lambda^{2 \alpha_3} - a_0}  .
\ee
Using Eq. (\ref{all-a}), we can write Eq. (\ref{w-G2}) 
in the form
\be \label{w-G3}
\hat w(x,\omega)= \frac{2 q_0}{E\, I}
\int^{\infty}_0 \frac{ \cos( \lambda |x|) \, d \lambda}{
 l^2_s (\alpha_2) \, \lambda^{2 \alpha_2} 
+ \lambda^{2 \alpha_1} 
- (\rho/E) \, l^2_d(\alpha_3) \, \omega^{2\beta}  \, \lambda^{2 \alpha_3}  
- (\rho \, A)/(E\, I) \, \omega^{2\beta} }  .
\ee
For a fractional non-local material without memory ($\beta=1$) 
and fractional acceleration gradient ($l^2_d(\alpha_3)=0$), 
the corresponding solution is
\be \label{w-G4}
\hat w(x,\omega)= \frac{2 q_0}{E\, I \,|x|}
\int^{\infty}_0 \frac{ \cos( \lambda |x|) \, d \lambda}{
 l^2_s (\alpha_2) \, \lambda^{2 \alpha_2} 
+ \lambda^{2 \alpha_1}  
- (\rho \, A)/(E\, I) \, \omega^{2} }  .
\ee
This is in fact, the solution given by Eq. (\ref{Solut-2}) 
for the fractional gradient Euler-Bernoulli beam equation 
of motion for the point load case of Eq. ({\ref{deltaf}}).

\subsection{Dispersion law and general harmonic solution of the combined \\ strain-acceleration fractional gradient beam model}

Let us now obtain a general harmonic solution of the combined strain-acceleration fractional gradient beam model
defined by Eq. (\ref{FEB-2com2}).
Using Property 2.34 in \cite{KST},
the Fourier transform of $(\, ^RD^{\alpha}_x w)(x)$ 
is given by Eq. (\ref{Four}),
where $w(x,t)$ belongs to the space $C^{\infty}_0(\mathbb{R}^2)$ 
of infinitely differentiable functions on 
$\mathbb{R}^2$ with a compact support.

The Fourier transform ${\cal F}$ of the fractional differential equation (\ref{FEB-2com2}) with $q=0$ gives
\be
-\rho A \,  |\omega|^{2\beta} 
+ E\, I \, |k|^{2\alpha_1} 
+ E\, I \, l^2_s (\alpha_2) \, |k|^{2\alpha_2} 
- \rho \, I \, l^2_d (\alpha_3) \, |k|^{2\alpha_3} \, |\omega|^{2\beta} =0 , 
\ee
implying that
\be
\omega^{2\beta} = \frac{E}{\rho} \, \frac{I}{A} \,
\frac{|k|^{2\alpha_1}+ l^2_s (\alpha_2) \, |k|^{2\alpha_2} }{1 + (I/A) \, l^2_d (\alpha_3) \, |k|^{2\alpha_3} } .
\ee
As a result, we obtain
\be \label{omega-beta}
\omega = C^{1/\beta}_e \, R^{1/\beta} \, |k|^{\alpha_1/\beta} \, \sqrt[2\beta]{ \frac{1+ l^2_s (\alpha_2) \, |k|^{2(\alpha_2-\alpha_1)} }{1 + R^2 l^2_d (\alpha_3) \, |k|^{2\alpha_3} } } ,
\ee
where
\be 
R = \sqrt{I/A} , \quad C_e = \sqrt{R / \rho} . \ee
The parameter $R$ is called the gyration radius.

In the absence of memory, i.e. $\beta=1$, 
equation (\ref{omega-beta}) yields 
\be \label{omega-beta=1}
\omega = C_e \, R \, |k|^{\alpha_1} \, \sqrt{ \frac{1+ l^2_s (\alpha_2) \, |k|^{2(\alpha_2-\alpha_1)} }{1 + R^2 l^2_d (\alpha_3) \, |k|^{2\alpha_3} } } .
\ee
If $\alpha_1=\alpha_3=2$, $\beta=1$ and $\alpha_2=3$,  
equation (\ref{omega-beta}) gives
\be \label{omega-beta=1b}
\omega = C_e \, R \, k^2 \, \sqrt{ \frac{1+ l^2_s \, k^2 }{1 + R^2 l^2_d \, k^4 } } ,
\ee
which is precisely the dispersion relation 
obtained earlier (Eq. (50) of \cite{AA2011})) for the non-fractional combined strain-acceleration gradient beam model.
Using Eq. (\ref{omega-beta}), we can obtain the group 
velocity $C_g =\partial \omega(k) / \partial k$ for 
the combined strain-acceleration 
fractional gradient beam model as
\[ \frac{C_g}{C_e} = \frac{1}{2\beta} C^{(1-\beta)/\beta}_e \, R^{1/\beta} \, \frac{ \left( 1+ l^2_s (\alpha_2) \, |k|^{2(\alpha_2-\alpha_1)} \right)^{(1-2\beta)/(2\beta)} }{ \left(1 + R^2 \, l^2_d (\alpha_3) \, |k|^{2\alpha_3} \right)^{(1+2\beta)/(2\beta)}}  
\cdot \Bigl( 2 \alpha_1 \, |k|^{2\alpha_1-1}+ 2\alpha_2 \, l^2_s (\alpha_2) \, |k|^{2\alpha_2-1} +
\]
\be \label{omega-group}
+ 2(\alpha_1+\alpha_3) \, R^2 \, l^2_d (\alpha_3) \, |k|^{2(\alpha_1 + \alpha_3)-1} 
+ 2(\alpha_2+\alpha_3) \, R^2 \, l^2_s (\alpha_2) \, l^2_d (\alpha_3) \, |k|^{2(\alpha_2 + \alpha_3)-1} \Bigr) .
\ee
If $\alpha_1=\alpha_3=2$, $\beta=1$ and $\alpha_2=3$, 
equation (\ref{omega-group}) is reduced to
\be \label{omega-group2}
 \frac{C_g}{C_e} = R \, k \, \frac{ 2 + 3 l^2_s k^2 + R^2 \, l^2_s \, l^2_d \, |k|^{6} }{(1+ l^2_s k^2)^{1/2} \, (1+ R^2 \, l^2_d k^4)^{3/2}} .
\ee
This is the well-known normalized form of the corresponding 
group velocity (see Eq. (51) of \cite{AA2011})
for non-fractional counterpart of the model.


\newpage
\section{Toward 3D fractional gradient elasticity}

To develop a fractional gradient elasticity theory 
in three-dimensions (3D), the following approaches may be used:

(1) An approach based on the Riesz fractional derivatives and 
integrals for $\mathbb{R}^3$ \cite{SKM,KST,Riesz}.
This approach is best suited for 3D problems
with spherical symmetry. 
The Riesz fractional derivative can be considered as a
non-integer power of the Laplacian.
Such a simple 3D fractional gradient elasticity model
based on the Riesz fractional derivatives 
has already been recently considered 
by the authors in \cite{TA2014}, and 
it can be naturally derived from 
lattice models with long-range interactions 
\cite{CEJP2013,ISRN2014,IJSS2014}.

(2) An approach based on fractional vector calculus. 
Currently, however fractional vector calculus is formulated 
for a Cartesian coordinate system only 
\cite{TarasovSpringer,AP2008}. 
The transformation from Cartesian to cylindrical, 
spherical or other coordinates
is prohibitively complicated for fractional derivatives.
It is connected with the fact that the formula of fractional derivative of a composite function 
(see Eq. 2.209 in Section 2.7.3 of \cite{Podlubny}) 
is very complex, i.e.,
\be
_aD^{\alpha}_x f(g(x)) = \frac{(x-a)^{\alpha}}{\Gamma(1-\alpha)} f(g(x)) 
+\sum^{\infty}_{k=1} C^{\alpha}_k \, \frac{k! (x-a)^{k-\alpha}}{\Gamma(k-\alpha+1)} \,
\sum^k_{m=1} (D^m_g f)(g(x)) \sum \prod^k_{r=1} \frac{1}{a_r!} \Bigl(\frac{(D^r_x g)(x)}{r!}\Bigr)^{a_r} ,
\ee
where $\sum$ extends over all combinations of non-negative integer 
values of $a_1$, $a_2$, . . . , $a_k$ such that 
\be
\sum^k_{r=1} r a_r=k , \quad \sum^k_{r} a_r=m.
\ee 

These two approaches which allow us to construct 
3D fractional nonlocal models of  
gradient elasticity are briefly discussed below.

\subsection{Fractional gradient elasticity based on Riesz derivative}

Three-dimensional fractional gradient elasticity models 
based on the Riesz fractional derivative are possible
due to the fact that this fractional derivative is a generalization of the Laplacian in $\mathbb{R}^n$
and, in fact, it can be considered as a non-integer power of
the Laplacian.
The corresponding 3D fractional gradient elasticity model 
is described by the following equation 
(for details see \cite{TA2014})
\be \label{FPDE-4}
c_{\alpha} \, ((- \, ^R\Delta)^{\alpha/2} u) ({\bf r}) + 
c_{\beta} \, ((-\, ^R\Delta)^{\beta/2} u) ({\bf r}) = f({\bf r})  \quad (\alpha> \beta),
\ee
where ${\bf r} \in \mathbb{R}^3$ and $r=|{\bf r}|$ are dimensionless variables, and $(- \, ^R\Delta)^{\alpha /2}$ is the Riesz fractional Laplacian of order $\alpha$ \cite{KST}. 
The coefficients ($c_{\alpha}$, $c_{\beta}$) are phenomenological constants and 
the rest of the symbols have their usual meaning,
with $u$ denoting the radial component of the displacement.

For $\alpha > 0$ and suitable functions $u({\bf r})$,
${\bf r} \in \mathbb{R}^3$, the Riesz fractional derivative
can be defined \cite{KST} in terms of the inverse Fourier transform ${\cal F}^{-1}$ by
\be
((- \, ^R\Delta)^{\alpha /2} u)({\bf r})= {\cal F}^{-1} \Bigl( |{\bf k}|^{\alpha} ({\cal F} u)({\bf k}) \Bigr) ,
\ee
where ${\bf k}$ denotes the wave vector, 
$\alpha > 0$ and $x \in \mathbb{R}^n$. 
The fractional Laplacian in the Riesz form is
usually defined in terms of the hypersingular integral 
\be \label{Def-L}
((-\Delta)^{\alpha/2}f)(x) =\frac{1}{d_n(m,\alpha)} \int_{\mathbb{R}^n} 
\frac{1}{|z|^{\alpha+n}} (\Delta^m_z f)(x) \, dz , 
\ee
where $m> \alpha>0$, and $(\Delta^m_z f)(x)$ 
is a finite difference of order $m$ of a function $f(x)$ with a vector step $z \in \mathbb{R}^n$ 
centered at the point $x \in \mathbb{R}^n$:
\[ (\Delta^m_z f)(x) =\sum^m_{k=0} (-1)^k \frac{m!}{k!(m-k)!}  \, f(x-kz) , \]
where the constant $d_n(m,\alpha)$ is defined by
\[ d_n(m,\alpha)=\frac{\pi^{1+n/2} A_m(\alpha)}{2^{\alpha} 
\Gamma(1+\alpha/2) \Gamma(n/2+\alpha/2) \sin (\pi \alpha/2)} ,  \]
with 
\[ A_m(\alpha)=\sum^m_{j=0} (-1)^{j-1} \frac{m!}{j!(m-j)!} \, j^{\alpha} . \]
The definition given by Eq. (\ref{Def-L}) 
for the fractional Laplacian of order $\alpha$ 
does not depend on the choice of $m>\alpha$.
Its Fourier transform ${\cal F}$ satisfies the
relationship 
$({\cal F} (-\Delta)^{\alpha/2} f)(k) = |k|^{\alpha} ({\cal F}f)(k)$, which is valid 
for the Lizorkin space \cite{SKM}
and the space $C^{\infty}(\mathbb{R}^n)$ of infinitely differentiable 
functions on $\mathbb{R}^n$ with compact support.

If $\alpha=4$ and $\beta=2$, we have the well-known equation 
of  gradient elasticity \cite{AA2011} for 
the non-fractional case, i.e.,
\be \label{GradEl}
c_2 \, \Delta u ({\bf r}) - c_4 \Delta^2 u ({\bf r}) + f({\bf r}) = 0,
\ee
where
\be \label{GradEl-2} 
c_2 = E , \quad c_4 = \pm \, l^2 \, E .
\ee

Equation (\ref{FPDE-4}) is a fractional partial differential 
equation with a particular solution (Section 5.5.1. of \cite{KST}) of the form
\be \label{phi-G4}
u({\bf r})= \int_{\mathbb{R}^3} 
G^3_{\alpha, \beta} ({\bf r} - {\bf r}^{\prime}) \, 
f({\bf r}^{\prime}) \, d^3 {\bf r}^{\prime},
\ee
where the Green's type function is given by the expression
\be \label{FGF2}
G^3_{\alpha, \beta}({\bf r})= 
\int_{\mathbb{R}^3} \frac{1}{c_{\alpha} |{\bf k}|^{\alpha} + c_{\beta} |{\bf k}|^{\beta}  } \
e^{ + i ({\bf k},{\bf r}) } \, d^3 {\bf k} .
\ee
Using Lemma 25.1 of \cite{SKM}, the kernel function 
in Eq. (\ref{FGF2}) can be represented by the equation
\be \label{G-4}
G^3_{\alpha, \beta} ({\bf r}) =\frac{1}{(2 \pi)^{3/2} \, \sqrt{|{\bf r}|}} 
\int^{\infty}_0 \frac{ \lambda^{3/2} \, J_{1/2} (\lambda |{\bf r}|) 
}{c_{\alpha} \lambda^{\alpha}+ c_{\beta} \lambda^{\beta}} 
\, d \lambda ,
\ee
where $J_{1/2}(z) = \sqrt{2/(\pi z)} \, \sin (z)$ denotes
Bessel function of the first kind.

If we consider the deformation of an infinite elastic continuum
due to an external field $f({\bf r})$ applied to 
a very small region, then for distances $|{\bf r}|$ 
which are large in comparison with the size of the region
(neighborhood) of load application, 
we can suppose that $f({\bf r})$ is applied at a point \cite{LL}:
\be \label{deltaf-2}
f({\bf r}) = f_0 \, \delta ({\bf r}) . 
\ee
Then, the displacement field $u ({\bf r})$  
has a simple form 
$u ({\bf r}) = f_0 \, G^3_{\alpha, \beta} ({\bf r})$
given by  
\be
u ({\bf r})
=\frac{1}{(2 \pi)^{3/2} \, \sqrt{|{\bf r}|}} 
\int^{\infty}_0 \frac{ \lambda^{3/2} \, J_{1/2} (\lambda |{\bf r}|) 
}{c_{\alpha} \lambda^{\alpha}+ c_{\beta} \lambda^{\beta}} 
\, d \lambda .
\ee



\subsection{Fractional vector calculus and 3D models}

\subsubsection{Fractional vector calculus}

Fractional vector calculus is a very important tool for
describing processes in complex media and materials with non-local properties. 
It allows us to formulate a dynamical theory of materials
with non-locality of power-law type in three dimensions.
At present, however, several formulations of fractional vector 
calculus are either incorrect or inconsistent, leading to errors. 
It seems that it is possible to define
a generalization of $\operatorname{grad}$, $\operatorname{div}$
and $\operatorname{curl}$ operators by using a fractional
derivative $D^{\alpha}_{x_k}$ instead of the usual 
derivative $D^1_{x_k}$, 
where $D^{\alpha}_{x_k}$ are fractional (Liouville, Riemann-Liouville, Caputo, etc.) 
derivatives of order $\alpha$ with respect to $x_k$, $k=1,2,3$.
In such an approach, there is considerable arbitrariness 
in the definition of vector operators.
The main problem in fractional vector calculus, however, appears 
when we try to generalize not only differential vector operators, 
but also the related integral theorems \cite{AP2008}.
In general, a robust framework of 
fractional vector calculus must include generalizations of
the differential operators (gradient, divergence, curl), 
the integral operations (flux, circulation), 
and the theorems of Gauss, Stokes and Green. 

The main problem in the formulation of fractional integral vector operations is connected with the complex form 
of the fractional analogue of the Newton-Leibniz formula
$\, _aI^1_b  \, _aD^1_x f(x) =f(b)- f(a)$.
In fact, the non-commutativity of $D^n_x$ and $_aI^{\alpha}_x$  
does not allows us to derive a convenient Riemann-Liouville
fractional counterpart of the Newton-Leibniz formula. 
For fractional Riemann-Liouville integrals and derivatives, 
we have the relation
\be \label{21}
_aI^{\alpha}_b  \, _aD^{\alpha}_x f (x) =f(b)-
\sum^{n}_{j=1} \frac{(b-a)^{\alpha-j}}{\Gamma(\alpha-j+1)}
(D^{n-j}_x \ _aI^{n-\alpha}_x f)(a) 
\ee
holding almost everywhere in $[a,b]$,
where $D^{n-j}_x=d^{n-j}/dx^{n-j}$ are integer derivatives,
and $n-1<\alpha<n$. 
Here $f(x)$ is a Lebesgue measurable function on $[a,b]$ for which $_aI^1_b f(x)< \infty$,
and $_aI^{n-\alpha}_b  f(x)$ has absolutely continuous derivatives
up to order $(n-1)$ on $[a,b]$. 
This relation was proved in \cite{SKM} 
(see Theorem 2.4 of Section 2.6).
For $0<\alpha<1$, Eq. (\ref{21}) gives
\be \label{22}
_aI^{\alpha}_b  \, _aD^{\alpha}_x f(x) =f(b)-
\frac{(b-a)^{\alpha-1}}{\Gamma(\alpha)} \ _aI^{1-\alpha}_b f(x) .
\ee
Obviously, that Eqs. (\ref{22}) and (\ref{21})
do not have the usual form of the Newton-Leibniz formula. 

A consistent formulation of fractional vector calculus has been realized in \cite{AP2008} by using fractional derivatives and fractional integrals of different types. 
For this purpose, the Riemann-Liouville integration and 
the Caputo differentiation are used.
The main property is that the Caputo fractional derivative  
provides an operation that is inverse to the Riemann-Liouville 
fractional integration from the left. 
As a result, we can formulate a fractional analogue 
of the Newton-Leibniz formula
in the usual form if the integral is of Riemann-Liouville type 
and the derivative is of the Caputo type. i.e.,
\be \label{FNLF}
 _aI^{\alpha}_b \ _a^CD^{\alpha}_x f(x)=f(b)-f(a) , \quad (0<\alpha<1) ,
\ee
where $ _a^CD^{\alpha}_x$ is the Caputo fractional derivative 
defined by the equation
\[ _a^CD^{\alpha}_x F(x) =\, _aI^{n-\alpha}_x D^n_x F(x)=
\frac{1}{\Gamma(n-\alpha)} \int^{x}_{a} 
\frac{dx' \, D^n_{x'} F(x')}{(x-x')^{1+\alpha-n}} , \]
where $n-1<\alpha <n$, and
$\ _aI^{\alpha}_x$ is the Riemann-Liouville fractional integral 
\[  _aI^{\alpha}_x  f(x) :=  \frac{1}{\Gamma(\alpha)} 
\int^{x}_{a} \frac{f(x')}{(x-x')^{1-\alpha}} dx' . \]
Here $f(x)$ is a real-valued function defined on a closed interval $[a, b]$
such that $f(x) \in AC^1[a,b]$ or $f(x)\in C^1[a,b]$.
For details, the reader may consult \cite{AP2008},
where the fractional differential operators 
are defined such that fractional generalizations of 
integral theorems (Green's, Stokes', Gauss') can be realized. 
Using this fractional vector calculus \cite{AP2008}, 
fractional differential equations for
the conservation of mass, momentum and energy
can be obtained for a continuum with power-law non-locality. 
This allows us to formulate 3D fractional models 
of continuum mechanics for fluids and solids with non-local properties.
In the next subsection, we show how the fractional vector 
calculus can be used to formulate a fractional generalization 
of gradient elasticity for the 3D case.

\subsubsection{Fractional differential vector operators}

To properly define fractional vector operations, we will first
introduce the operators that correspond to fractional 
differentiation and fractional integration. 
The left-sided Riemann-Liouville fractional integral operator 
is defined as
\be \label{operator-RLI}
_aI^{\alpha}_x[x']:=\frac{1}{\Gamma(\alpha)} 
\int^x_a \frac{dx'}{(x-x')^{1-\alpha}} , \quad (\alpha >0) .
\ee
To designate that the operator given by Eq. (\ref{operator-RLI}) 
acts on a real-valued function $f(x)\in L_1[a,b]$,  
we employ the notation $_aI^{\alpha}_x[x']f(x')$.
We define the left-sided Caputo fractional differential operator
on $[a,b]$ in the form
\be \label{operator-Caputo}
_a^CD^{\alpha}_x[x']:=
\frac{1}{\Gamma(n-\alpha)} \int^x_a  \frac{dx'}{(x-x')^{1+\alpha-n}} 
\frac{\partial^n}{\partial {x'}^n} , \quad (n-1<\alpha<n) .
\ee
The Caputo operator defined by Eq. (\ref{operator-Caputo}) 
acts on real-valued 
functions $f(x)\in AC^n[a,b]$ as $_a^CD^{\alpha}_x[x']f(x')$.
We note that the Caputo operator can be represented as 
\[ 
_a^CD^{\alpha}_x[x']= _aI^{n-\alpha}_x [x'] D^n[x'] , 
\quad (n-1<\alpha < n) . 
\]
Equation (\ref{FNLF}) can be rewritten in the form
\be \label{FNLF2}
_aI^{\alpha}_b[x] \ _a^CD^{\alpha}_x[x'] f(x') = f(b)-f(a) , \quad (0<\alpha<1) . \ee

In the notations $_aI^{\alpha}_b[x]$ and $_aD^{\alpha}_x[x']$,
we idicate the variable of integration by the brackets $[ \, ]$,
and the lower indices show the limits of integration.
Note that in Eq. (\ref{FNLF2}) the variable of integration 
is $x$ since the result of the integration 
with respect to $x'$ in the operator 
$ _a^CD^{\alpha}_x[x']$ depends on $x$ only.
These notations are more convenient than the ones
usually used (see Eq.(\ref{FNLF})),
since it allows us to take into account the variables of 
integration and the domain of the operators.


We define a fractional generalization of nabla operator by
\be \label{fnabla}
\nabla^{\alpha}_W=\, ^C{\bf D}^{\alpha}_W = {\bf e}_1 \, ^CD^{\alpha}_W [x]+
{\bf e}_2 \, ^CD^{\alpha}_W [y]+{\bf e}_3 \, ^CD^{\alpha}_W [z] , 
\quad (n-1<\alpha < n) ,
\ee
where $\, ^CD^{\alpha}_{W}[x_m]$ denotes the Caputo fractional 
derivative with respect to coordinates $x_m$.
For the parallelepiped 
$W:=\{ a \leqslant x \leqslant b, 
\quad c \leqslant y \leqslant d, \quad g \leqslant z \leqslant h \}$, we have
\[ ^CD^{\alpha}_W [x] =\, _a^CD^{\alpha}_b [x] , \quad
^CD^{\alpha}_W [y]=\, _c^CD^{\alpha}_d [y] , \quad
^CD^{\alpha}_W [z]=\, _g^CD^{\alpha}_h [z] . \]

Let us now give the definitions of fractional 
gradient, divergence and curl operators 
in Cartesian coordinates \cite{TarasovSpringer,AP2008}).
We assume that $f(x)$ and ${\bf F}(x)$ are real-valued functions 
with continuous derivatives up to order $(n-1)$ on 
$W \subset \mathbb{R}^3$,   
such that their $(n-1)$ derivatives are absolutely continuous, 
i.e., $f,{\bf F} \in AC^n[W]$. 

(1) The fractional gradient is defined by 
\[
\operatorname{Grad}^{\alpha}_W f=\,  ^C{\bf D}^{\alpha}_W f=
 {\bf e}_l \ ^CD^{\alpha}_W[x_l] f(x,y,z)=
\]
\be
= {\bf e}_1 \ ^CD^{\alpha}_W[x] f(x,y,z)+ {\bf e}_2  
\ ^CD^{\alpha}_W[y] f(x,y,z)+  {\bf e}_3 \ ^CD^{\alpha}_W[z] f(x,y,z) ,
\ee
where $f=f(x,y,z)$ is a $(n-1)$ times continuously differentiable 
scalar field such that the derivative 
$D^{n-1}_{x_l} f$ is absolutely continuous.

(2) The fractional divergence is defined by the equation
\[
\operatorname{Div}^{\alpha}_W {\bf F}= \Bigl(\, ^C{\bf D}^{\alpha}_W , {\bf F}\Bigr)=
\, ^CD^{\alpha}_W[x_l] F_l(x,y,z) =
\]
\be
=\, ^CD^{\alpha}_W[x] F_x(x,y,z)+
\, ^CD^{\alpha}_W[y] F_y(x,y,z)+ \ ^CD^{\alpha}_W[z] F_z(x,y,z) ,
\ee
where ${\bf F}(x,y,z)$ is a $(n-1)$ times continuously differentiable 
vector field such that the derivatives $D^{n-1}_{x_l} F_l$ are absolutely continuous.

(3) The fractional curl operator is defined by
\[
\operatorname{Curl}^{\alpha}_W {\bf F}=\Bigl[\, ^C{\bf D}^{\alpha}_W, {\bf F} \Bigr]=
{\bf e}_l \, \varepsilon_{lmk} \, ^CD^{\alpha}_{W} [x_m] F_k= 
{\bf e}_1 \left(\ ^CD^{\alpha}_{W} [y] F_z-\ ^CD^{\alpha}_{W} [z] F_y \right) +
\]
\be
+{\bf e}_2  \left(\ ^CD^{\alpha}_{W} [z] F_x-\ ^CD^{\alpha}_{W} [x] F_z \right) 
+{\bf e}_3  \left(\ ^CD^{\alpha}_{W} [x] F_y-\ ^CD^{\alpha}_{W} [y] F_x \right) ,
\ee
where $F_k=F_k(x,y,z)\in AC^n[W]$, $(k=1,2,3)$.

(4) Using the notation introduced in Eq. (\ref{fnabla}),
the operator $(\, ^C{\bf D}^{\alpha}_W)^2 $ can be considered as the fractional Laplacian of the Caputo type:
\be \label{Delta-C}
^C\Delta^{\alpha}_W = \Bigl(\, ^C{\bf D}^{\alpha}_W,\, ^C{\bf D}^{\alpha}_W \Bigr) 
=(\, ^C{\bf D}^{\alpha}_W )^2 = \sum^3_{l=1}( ^CD^{\alpha}_W [x_l])^2.
\ee
Note that in the general case we have the inequality
\be
(\, ^CD^{\alpha}_W [x_l])^2 \ne \, ^CD^{2 \alpha}_W [x_l] .
\ee

Let us now give the basic relations for the fractional 
differential vector operators 
(for details of proofs see \cite{TarasovSpringer,AP2008}). 

(i) For the scalar field $f=f(x,y,z)$, we have
\be
\operatorname{Div}^{\alpha}_W \, \operatorname{Grad}^{\alpha}_W f =
\ ^CD^{\alpha}_W [x_l] \ ^CD^{\alpha}_W[x_l] f  =
\sum^3_{l=1}( ^CD^{\alpha}_W [x_l])^2 f .
\ee
Using then the notations introduced in Eqs. 
(\ref{fnabla}) and (\ref{Delta-C}), we conclude
\be
\operatorname{Div}^{\alpha}_W \, \operatorname{Grad}^{\alpha}_W   
=\Bigl(\, ^C{\bf D}^{\alpha}_W,\, ^C{\bf D}^{\alpha}_W \Bigr)
= \, ^C\Delta^{\alpha}_W .
\ee

(ii)
The second relation for the scalar field $f=f(x,y,z)$ is
\be
\operatorname{Curl}^{\alpha}_W \, \operatorname{Grad}^{\alpha}_W f =0 .
\ee

(iii)
For the vector field ${\bf F}={\bf e}_m F_m$, 
it is easy to prove the relation
\be
\operatorname{Div}^{\alpha}_W \, \operatorname{Curl}^{\alpha}_W {\bf F}(x,y,z) =0 .
\ee

(iv)
The following identity also holds for the double curl operator 
\be \label{CC}
\operatorname{Curl}^{\alpha}_W \, \operatorname{Curl}^{\alpha}_W \, {\bf F} = 
\operatorname{Grad}^{\alpha}_W \, \operatorname{Div}^{\alpha}_W \, {\bf F} -
(\, ^C{\bf D}^{\alpha}_W)^2 {\bf F} .
\ee

(v)
The Leibniz rule for fractional differential vector operators \cite{CNSNS2013} does not hold, i.e.,
\be
\operatorname{Grad}^{\alpha}_W \Bigl( f g \Bigr) \ne
\Bigl( \operatorname{Grad}^{\alpha}_W f \Bigr) g+
\Bigl( \operatorname{Grad}^{\alpha}_W g \Bigr) f ,
\ee
\be
\operatorname{Div}^{\alpha}_W \Bigl( f {\bf F} \Bigr) \ne
\Bigl( \operatorname{Grad}^{\alpha}_W f , {\bf F} \Bigr)+
f \ \operatorname{Div}^{\alpha}_W {\bf F} .
\ee

We define the fractional differential vector operators 
such that the fractional vector integral operators 
(circulation, flux, and volume integral) exist
as inverse operations.
This allows us to establish the fractional analogues 
of Green's, Stokes' and Gauss' integral theorems  
\cite{TarasovSpringer,AP2008}.
It is also noted that the fractional differential operators are nonlocal by definition.
The fractional gradient, divergence and curl operators
depend on the region $W$. 
This property allows for the use of fractional vector calculus
to describe complex materials with power-law non-locality 
in three dimensional space.
Note that these continuum fractional vector operators 
can be connected with the fractional-order operators 
on lattices with long-range interactions \cite{JPA2014}.

\subsection{Fractional 3D gradient elasticity model}

The simplest form of the stress-strain relation 
of gradient elasticity theory
can be written \cite{AA2011} as
\be \label{CE-1}
\sigma_{ij} = C_{ijkl} \Bigl( \varepsilon_{kl} \pm l^2_s \, \Delta \varepsilon_{kl} \Bigr) ,
\ee
where $C_{ijkl}$ is the matrix of elastic modulus,
$l_s$ is a length scale parameter, 
$\sigma_{ij}$ is the stress, and
$\varepsilon_{kl}$ is the strain tensor.
For homogenous and isotropic materials we have
\be
C_{ijkl} = \lambda \, \delta_{ij} \delta_{kl} + 2 \mu \, \delta_{ik} \delta_{jl} ,
\ee
where $\lambda$ and $\mu$ are the usual Lame constants, 
and $\delta_{ij}$ is the Kronecker delta.

The equation of motion based on Eq. (\ref{CE-1}) has the form
\be \label{EM-1}
C_{ijkl} \Bigl( D^1_{x_j} D^1_{x_l} \pm 
l^2_s D^1_{x_j} (D^1_{x_m} D^1_{x_m}) D^1_{x_l}  \Bigr) \, u_k +f_i = \rho \, D^2_t u_i ,
\ee
where $f_i$ are the components of the external force field,
and $u_k$ are the components of the displacement vector field.
For homogenous and isotropic materials, Eq. (\ref{EM-1}) 
can be written as
\be \label{EM-2}
\lambda \, \Bigl( D^1_{x_i} D^1_{x_k} \pm 
l^2_s D^1_{x_i} D^1_{x_k} (D^1_{x_m} D^1_{x_m}) \Bigr) \, u_k
 + 2 \mu \, \Bigl( (D^1_{x_l} D^1_{x_l}) \pm 
l^2_s (D^1_{x_l} D^1_{x_l}) (D^2_{x_m} D^1_{x_m})  \Bigr) \, u_i +f_i = \rho \, D^2_t u_i .
\ee
Using now operations of the vector calculus operators,
this equation can be rewritten in the following vector form 
\be \label{EM-3}
\lambda \, \Bigl( 1 \pm l^2_s \Delta \Bigr) \, \operatorname{grad} \operatorname{div} {\bf u}
 + 2 \mu \, \Bigl( \Delta \pm l^2_s \Delta^2 \Bigr) \, {\bf u} + {\bf f} = \rho \, D^2_t {\bf u} .
\ee


A formal fractional generalization of Eq. (\ref{EM-1}) 
can be obtained in the form
\be \label{EM-0f}
C_{ijkl} \Bigl( \, ^CD^{\alpha_j}_W [x_j] \, ^CD^{\alpha_l}_W [x_l] \pm 
l^2_s(\alpha) \, ^CD^{\alpha_j}_W [x_j] (\, ^CD^{\alpha_m}_W [x_m] \, ^CD^{\alpha_m}_W [x_m]) \, ^CD^{\alpha_l}_W [x_l]  \Bigr) \, u_k +f_i = \rho \, D^2_t u_i ,
\ee
where $\alpha=(\alpha_1,\alpha_2,\alpha_3)$ is a multi-index.
For the isotropic case ($\alpha_1=\alpha_2=\alpha_3=\alpha$), 
we have the fractional equation 
\be \label{EM-1f}
C_{ijkl} \Bigl( \, ^CD^{\alpha}_W [x_j] \, ^CD^{\alpha}_W [x_l] \pm 
l^2_s(\alpha) \, ^CD^{\alpha}_W [x_j] (\, ^CD^{\alpha}_W [x_m] \, ^CD^{\alpha}_W [x_m]) \, ^CD^{\alpha}_W [x_l]  \Bigr) \, u_k +f_i 
= \rho \, D^2_t u_i .
\ee
Using the properties of the fractional 
differential vector operators, 
Eq. (\ref{EM-1f}) for homogenous and isotropic materials 
can be rewritten in the following vector form 
\be \label{EM-2f}
\lambda \, \Bigl( 1 \pm l^2_s(\alpha) \, ^C\Delta^{\alpha}_W  \Bigr) \, \operatorname{Grad}^{\alpha}_W \operatorname{Div}^{\alpha} {\bf u}
 + 2 \mu \, \Bigl( \, ^C\Delta^{\alpha}_W  \pm l^2_s(\alpha) (\, ^C\Delta^{\alpha}_W)^2  \Bigr) \, {\bf u} + {\bf f} = 
\rho \, D^2_t {\bf u} .
\ee
Note that, in general, the following inequality holds
\be
(\, ^C\Delta^{\alpha}_W)^2 \ne \, ^C\Delta^{2\alpha}_W ,
\ee
since $(\, ^CD^{\alpha}_x)^2 \ne \, ^CD^{2\alpha}_x$.

In general, the fractional equations of motion may 
contain expressions of the form
$A_{\alpha}(x_j) \, ^CD^{\alpha}_W [x_j]$ 
with a given function $A_{\alpha}(x)$ 
instead of the fractional derivative $\, ^CD^{\alpha}_W [x_j]$.  
The explicit form of the function $A_{\alpha}(x)$ is deduced 
by the conservation law for non-local media by using 
the fractional vector calculus \cite{TarasovSpringer,AP2008}.
In this case, the resulting 3D gradient elasticity models 
are more complicated and the corresponding equations 
of motion are much more difficult to solve. 
To solve the governing equations of 3D fractional models
we should also take into account an explicit form of the violation of 
the semigroup property for the Caputo derivative \cite{TA2014}
that gives the relationship between 
the product $\, ^CD^{\alpha}_{a+} \, ^CD^{\beta}_{a+} $ and 
the derivative $\, ^CD^{\alpha+\beta}_{a+}$.


Using Eq. (\ref{CC}) in the form  
\be \label{CC2}
\operatorname{Grad}^{\alpha}_W \, \operatorname{Div}^{\alpha}_W \, {\bf u} =
\operatorname{Curl}^{\alpha}_W \, \operatorname{Curl}^{\alpha}_W \, {\bf u} + 
\, ^C\Delta_W  {\bf u} ,
\ee
we can rewrite Eq. (\ref{EM-2f}) as
\be \label{EM-2f-new}
\lambda \, \Bigl( 1 \pm l^2_s(\alpha) \, ^C\Delta^{\alpha}_W  \Bigr) \, \operatorname{Curl}^{\alpha}_W \, \operatorname{Curl}^{\alpha}_W \, {\bf u}
 + (\lambda+2 \mu ) \, \Bigl( \, ^C\Delta^{\alpha}_W  \pm l^2_s(\alpha) (\, ^C\Delta^{\alpha}_W)^2  \Bigr) \, {\bf u} + {\bf f} = \rho \, D^2_t {\bf u} .
\ee
If we futher assume that the displacement vector ${\bf u}$
is radial and function of $r=|{\bf r}|$ 
alone ($u_k=u_k(|{\bf r}|)$), we have 
\[ \operatorname{Curl}^{\alpha}_W \, {\bf u}=0 , \] 
and, as a result, Eq. (\ref{EM-2f-new}) has the form
\be \label{EM-2f-new2}
(\lambda+2 \mu ) \, \Bigl( 1 \pm l^2_s(\alpha) \, ^C\Delta^{\alpha}_W \Bigr) 
\, ^C\Delta^{\alpha}_W  \,{\bf u} +{\bf f} 
= \rho \, D^2_t {\bf u} .
\ee
This is the governing fractional gradient elasticity equation 
for homogenous and isotropic materials with spherical symmetry.


\subsection{The square of fractional derivative is not equal to a dual-order derivative}

In order to solve the governing equations of 
fractional gradient elasticity,
we should give first the explicit form of the relationship 
between the square of the Caputo derivative 
$(\, ^CD^{\alpha}_{a+})^2$ and 
the Caputo derivative $\, ^CD^{2\alpha}_{a+}$.
To obtain this relation we use Eq. 2.4.6 of \cite{KST}, 
in the form
\be
(\, ^CD^{\alpha}_{a+} f)(x)\, = (\, ^{RL}D^{\alpha}_{a+} f)(x) -
\sum^{n-1}_{k=0} \frac{(D^k f)(a)}{\Gamma(k-\alpha+1)} (x-a)^{k-\alpha} ,
\ee
and Eq. 2.1.16 of \cite{KST}, in the form
\be \label{Int-a}
I^{\alpha}_{a+} (x-a)^{\beta} = 
\frac{\Gamma(\beta+1)}{\Gamma(\alpha+\beta)} (x-a)^{\beta+\alpha} ,
\ee
where $\alpha>0$ and $\beta>-1$.
The condition $\beta>-1$ gives another 
restriction for $\alpha$ in the form $\alpha<1$.
The relationship between the square of the Caputo derivative of order $\alpha$
and the Caputo derivative of order $2\alpha$ takes then the form
\be \label{Ca2}
(\, ^CD^{\alpha}_{a+})^2 f(x) =
\, ^CD^{2\alpha}_{a+} f(x) +
\frac{f^{\prime}(a)}{\Gamma(1-2\alpha)} 
(x-a)^{1-2\alpha} , \quad (0<\alpha \le 1) ,
\ee
where $\alpha \ne 1/2$.
Using Eq. (\ref{Ca2}), we can represent the fractional Laplacian of Caputo type as
\be \label{2Lap-Lap2}
\, ^C\Delta^{\alpha}_W = \sum^3_{k=1} \, ^CD^{2\alpha}_{x_i}
+  \sum^3_{k=1} 
\frac{(D^1_{x_k}f)(a)}{\Gamma(1-2\alpha)} (x_k-a_k)^{1-2\alpha} .
\ee

Note that the relation given by Eq. (\ref{Ca2}) 
cannot be used for $\alpha>1$.
As a result, additional difficulties for solving fractional gradient equations arise.
To solve these problems, we can use a generalization of the
Ru-Aifantis operator split method \cite{RA1993,AMA2008}.

\subsection{Operator split method for fractional gradient elasticity}


In 1993, Ru and Aifantis \cite{RA1993} suggested an operator 
split method to solve static problems of gradient elasticity.
Let us consider a generalization of this method to solve 
the fractional gradient elasticity problems. For the static case, 
Eq. (\ref{EM-2f}) can be written in the form
\be \label{EM-2f-2}
\Bigl( 1 \pm l^2_s(\alpha) \, ^C\Delta^{\alpha}_W  \Bigr) \, \Bigl[ \lambda \,  \operatorname{Grad}^{\alpha}_W \operatorname{Div}^{\alpha} + 2 \mu \, ^C\Delta^{\alpha}_W \Bigr] \, {\bf u}  + {\bf f} = 0 .
\ee
By introducing $l^2_s(\alpha)=0$ in Eq. (\ref{EM-2f-2}), 
we obtain the fractional differential equation 
\be \label{A-2}
L^{(\alpha)} \, {\bf u} + {\bf f}=0 , 
\ee
where we use the fractional operator
\be \label{La}
L^{(\alpha)} =  \lambda \,  \operatorname{Grad}^{\alpha}_W \operatorname{Div}^{\alpha} + 2 \mu \, ^C\Delta^{\alpha}_W .
\ee
For the gradient-dependent case $l^2_s(\alpha) \ne 0$, 
Eq. (\ref{EM-2f-2}) has the form
\be \label{A-3}
\Bigl( 1 \pm l^2_s(\alpha) \, ^C\Delta^{\alpha}_W  \Bigr)
\, L^{(\alpha)} \, {\bf u} +{\bf f} = 0 .
\ee

In general, it is necessary to solve the fractional partial differential equation of order $4\alpha$, which has 
a very complex form caused by the inequality
$(\, ^C\Delta^{\alpha}_W )^2 \ne 
\, ^C\Delta^{2\alpha}_W$ for the 
fractional Laplacian of Caputo type.
The following observation can reduce the complexity of 
this task and greatly facilitate
the obtaining of solutions in certain cases.
For the radial displacement case
($\operatorname{Curl}^{\alpha}_W \, {\bf u}=0 $),
the operators $L^{(\alpha)}$ and $\, ^C\Delta^{\alpha}_W $ commute, i.e.,
\[ L^{(\alpha)} \, \, ^C\Delta^{\alpha}_W - \, ^C\Delta^{\alpha}_W   \, L^{(\alpha)} =0 . \]
Therefore, we can see from Eq. (\ref{A-3}) that the vector field $\Bigl( 1 \pm l^2_s(\alpha) \, ^C\Delta^{\alpha}_W  \Bigr) \, {\bf u}$
satisfies the non-gradient expression of Eq. (\ref{A-2}) 
for the field ${\bf u}$.
Thus, if $\Bigl( 1 \pm l^2_s(\alpha) \, ^C\Delta^{\alpha}_W  \Bigr) \, {\bf u}$
can be identified with the non-gradient displacement field ${\bf u}^c$ of fractional non-gradient elasticity theory
given by Eq. (\ref{A-2}), which
can be solved, then the original fractional gradient elasticity theory given by Eq. (\ref{EM-2f-2})
is reduced to the following fractional equation
\be \label{A-4}
\Bigl( 1 \pm l^2_s(\alpha) \, ^C\Delta^{\alpha}_W  \Bigr) \, {\bf u}^g={\bf u}^c ,
\ee
where ${\bf u}^c$ is a classical ("non-gradient") solution of the fractional equation
\[ L^{(\alpha)} \, {\bf u}^c +{\bf f}=0 . \]
Obviously, the solution of Eq. (\ref{A-4}) can be more 
conveniently obtained.
This establishes a connection between the "gradient" (g) and the non-gradient  "classical" (c) fractional elasticity solutions. 


For the non-radial case ($\operatorname{Curl}^{\alpha}_W \, {\bf u} \ne 0 $), the fractional gradient elasticity theory
given by Eq. (\ref{EM-1f}) takes the form
\be \label{12}
L^{(\alpha)}_{ik} \, \Bigl( 1 \pm l^2_s(\alpha) \, ^C\Delta^{\alpha}_W  \Bigr) \, u_k + f_i=0 ,
\ee
where
\be
L^{(\alpha)}_{ik} = C_{ijkl} \, ^CD^{\alpha}_W [x_j] \, ^CD^{\alpha}_W [x_l] .
\ee
Using the operator split approach, Eq. (\ref{12}) can be solved 
as an uncoupled sequence of two sets of fractional equations, that is
\be \label{13}
L^{(\alpha)}_{ik} \, u^c_k + f_i=0 
\ee
followed by 
\be \label{14}
\Bigl( 1 \pm l^2_s(\alpha) \, ^C\Delta^{\alpha}_W  \Bigr) \, u^g_k = u^c_k  ,
\ee
where two separate displacement fields are distinguished. 
Firstly, $u^c_k$ obeys the non-gradient fractional elasticity 
as given by Eq. (\ref{13}). 
Secondly, $u^g_k$ are the same as $u_k$ in Eq. (\ref{12}), 
but they are now appended with a superscript $g$ to emphasize 
that they incorporate fractional gradient effects.

\subsection{Solutions by fractional operator split method}

Unfortunately, the applicability of
fractional vector calculus to solve 3D 
fractional differential equations, such as Eq. (\ref{EM-2f}),
is very limited due to the weak development of this area of mathematics.
Therefore, we demonstrate an application of 
the suggested generalization of the operator split method 
to obtain solutions of fractional gradient elasticity equation for a 1D case only.
The 1D counterpart of Eq. (\ref{EM-2f-new2}) reads
\be \label{EM-2f-new3}
(\lambda+2 \mu ) \, \Bigl( 1 \pm l^2_s(\alpha) 
\, ^C\Delta^{\alpha}_x \Bigr) \,
\, ^C\Delta^{\alpha}_x \, u(x) + f(x)=0.
\ee
Using the operator split method in Eq. (\ref{EM-2f-new3}), 
we derive two uncoupled fractional equations
\be \label{Ru-1D-1}
L^{(\alpha)} \, u^c(x) + f(x)=0  ,
\ee
and  
\be \label{Ru-1D-2}
\Bigl( 1 \pm l^2_s(\alpha) \, ^C\Delta^{\alpha}_x \Bigr) 
\, u^g (x) = u^c (x) ,
\ee
where the notation
$L^{(\alpha)} = 
(\lambda+2 \mu ) \, ^C\Delta^{\alpha}_x$ was used. 

Let us first consider the equation for the non-gradient case.
Using (\ref{Ca2}), equation (\ref{Ru-1D-1}) can be represented as
\be \label{Ru-1D-3}
(\lambda+2 \mu ) \, ^CD^{2\alpha}_{a+} u(x) +
\frac{(\lambda+2 \mu ) \, u^{\prime}(a)}{\Gamma(1-2\alpha)} 
(x-a)^{1-2\alpha} + f(x)=0 .
\ee
We can rewrite this equation in the form
\be \label{Ru-1D-4}
(\lambda+2 \mu ) \, ^CD^{2\alpha}_{a+} u(x) + f_{eff}(x) =0 ,
\ee
where we have used the effective body force given 
by the expression
\be
f_{eff}(x) =  \frac{(\lambda+2 \mu )\, u^{\prime}(a)}{\Gamma(1-2\alpha)} 
(x-a)^{1-2\alpha} + f(x) .
\ee
If $f_{eff}(x) \in C_{\gamma}[a;b]$ with  
$0 \le \gamma <1$ and $\gamma \le 2 \alpha$, then
(see Section 4.1.3 and Theorem 4.3 of \cite{KST})
equation (\ref{Ru-1D-4}) has a unique solution $u^c(x)$ 
belonging to the space $C^{2\alpha,n}_{\gamma}[a;b]$, 
where $n-1 < 2\alpha <n$, defined by the expression
\be \label{uc}
u(x)= u^c(x) = \sum^{n-1}_{k=0} \frac{u^{(k)}(a)}{k!} \, (x-a)^k 
- \frac{1}{(\lambda+2 \mu ) \, \Gamma (2\alpha)} 
\int^{x}_{a} \frac{f_{eff}(z)}{(x-z)^{1-2\alpha}} ,
\ee
where $n-1 < 2\alpha <n$.


Next, we consider the corresponding  equation for 
the gradient case.
Equation (\ref{Ru-1D-2}) can be rewritten as
\be \label{Ru-1D-5}
^CD^{2\alpha}_{a+} \, u^g (x) 
\pm \, l^{-2}_s(\alpha)  \, u^g (x)  = \pm \, u^c (x) .
\ee
Using (\ref{Ca2}), equation (\ref{Ru-1D-5}) 
can be represented as
\be \label{Ru-1D-6}
^CD^{2\alpha}_{a+} \, u^g (x) 
+ \frac{u^{\prime}(a)}{\Gamma(1-2\alpha)} (x-a)^{1-2\alpha} 
\pm \, l^{-2}_s(\alpha) \, u^g (x)  = \pm \, u^c (x) ,
\ee
where $u^c(x)$ is defined by Eq. (\ref{uc}).
We rewrite this equation in the form
\be \label{Ru-1D-7}
^C\Delta^{3\alpha}_x \, u^g (x) 
\pm \, l^{-2}_s(\alpha) \, u^g (x)  = u^c_{eff} (x) ,
\ee
\be
u^c_{eff}(x) = \pm u^c (x) - 
\frac{u^{\prime}(a)}{\Gamma(1-2\alpha)} (x-a)^{1-2\alpha} .
\ee
If $u^c_{eff}(x) \in C_{\gamma}[a;b]$ with  
$0 \le \gamma <1$ and $\gamma \le 2 \alpha$, then
(see Theorem 4.3 of \cite{KST})
equation (\ref{Ru-1D-7}) has a unique solution $u^g(x)$ 
belonging to the space $C^{2\alpha,n}_{\gamma}[a;b]$, 
where $n-1 < 2\alpha <n$, defined by the expression
\[
u^g(x) = \sum^{n-1}_{k=0} u^g(a) \, (x-a)^{k} \,
E_{2\alpha, k+1} [ \mp \, l^{-2}_s(\alpha) \, (x-a)^{2\alpha}] + \]
\be
+\int^{x}_{a} (x-z)^{2\alpha -1} \, E_{2\alpha,2 \alpha}
[\pm \, l^{-2}_s(\alpha) \, (x-z)^{2\alpha}] \, u^c_{eff}(z) \, dz . 
\ee
The quantity $E_{\alpha,\beta}(z)$ is the Mittag-Leffler 
function \cite{KST} defined by the relation
\be \label{Eab}
E_{\alpha,\beta}[z] = \sum^{\infty}_{k=0} \frac{z^k}{\Gamma(\alpha k+\beta)} , 
\quad (\alpha>0, \beta \in \mathbb{R}) .
\ee
Note also that $E_{1,1}[z]=e^z$.




\newpage
\section{Toward gradient elasticity of fractal materials}

Fractals are measurable metric sets with non-integer 
Hausdorff dimension \cite{Fractal1,Fractal2} 
that should be observed on all scales. 
Real fractal materials can be characterized by 
an asymptotic relation between
the mass $M(W)$ and the volume $V(W)$
of regions $W$ of the fractal medium.
For example, for a homogeneous fractal medium,
a ball of radius $R \gg R_0$ contains 
the mass $M_D(W) =M_0 (R/ R_0)^D$, 
where the number $D$ is called the mass dimension,
and $R_0$ is a characteristic size 
related to the arrangement of the medium particles. 
The mass dimension $D$ does not depend on 
the shape of the region $W$,
or on the packing of particles (close packing,
random packing or porous packing with uniform distribution of holes).

As a result, we can define a fractal material as a medium 
with non-integer mass (or number of particles) dimension. 
Although, the non-integer dimension does not reflect 
completely the geometric and dynamic properties of 
a fractal medium, it nevertheless permits a number of 
important conclusions about its behavior.


\subsection{Fractional continuum model for fractal materials}

In general, a fractal material cannot be considered 
as a usual continuum, since there are places and areas 
that are not filled with particles.
Nevertheless it can be described by special continuum models 
\cite{PLA2005-1,AP2005-2,TarasovSpringer} based
on the use of the integrals with non-integer order.
The order of these integrals should be defined by 
the fractal mass dimension.
The kernel of the fractional integral operator describes a density of permitted states (permitted places) in space.
The fractional-order integrals can be considered 
as integrals over a non-integer dimensional space 
up to a numerical factor by using the well-known 
formulas of dimensional regularization \cite{Collins}.  

Fractional integral continuum models of fractal media  
may have a wide range of applications \cite{TarasovSpringer}
due to the relatively small numbers of parameters 
that define a fractal material of great complexity 
and rich structure.
One of the advantages of such models is the ability 
to describe dynamics of fractal materials and media 
(for details see \cite{TarasovSpringer}). 

To describe fractal materials by a fractional integral 
continuum model, we use two different notions: 
the density of states $c_n(D,{\bf r})$
and the distribution function $\rho({\bf r},t)$.

(1) The function $c_n(D,{\bf r})$ is a density of states 
in the $n$-dimensional Euclidean space $\mathbb{R}^n$.
The density of states describes
how closely packed permitted states of particles 
in the space $\mathbb{R}^n$.
The expression $c_n(D,{\bf r})\, dV_n$
represents the number of states (permitted places) between $V_n$ and $V_n +dV_n$.
We note that the symmetry of the density of states 
$c_n(D,{\bf r})$ must be the defined by the symmetry properties 
of the fractal medium.

(2) The function $\rho({\bf r},t)$ is a distribution function
in the $n$-dimensional Euclidean space $\mathbb{R}^n$.
It describes the distribution of physical values
(for example, mass, electric charge, number of particles, probability)
on a set of possible (permitted) states 
in the space $\mathbb{R}^n$.


For example, the mass of a region $dV_n$ in fractal media
is defined by the equation
\[ dM({\bf r},t)= \rho({\bf r},t) c_n(D,{\bf r}) dV_n . \]
In general, we cannot consider the value 
$\rho({\bf r},t) c_n(D,{\bf r})$
as a new distribution function or a particle number density,
since the notions of density of states and 
of distribution function are different.
We cannot reduce all properties of the system 
to a description of the distribution function.
This fact is well-known in statistical and 
condensed matter physics, where the density of states 
is usually considered as a density of energy states or 
as a density of wave vector states \cite{Kittel}
that describe how closely packed 
the allowed states in energy or wave-vector spaces.
For fractal distributions of particles in a coordinate space $\mathbb{R}^n$, we must use a density of states in this space.
The density of states $c_n(D,{\bf r})$ in $\mathbb{R}^n$
is chosen such that
$d \mu_D({\bf r},n) = c_n(D,{\bf r})dV_n$
describes the number of states in $dV_n$.
We use the notations 
\[ dV_D=c_3(D,{\bf r})dV_3 , \quad 
dS_d=c_2(d,{\bf r})dS_2 , \quad 
dl_{\beta}=c_1(\beta,{\bf r})dl_1 \]
to describe densities of states in 
$n$-dimensional Euclidean spaces with $n=1,2,3$.

\subsection{Mass of fractal materials}

The cornerstone of fractal media is the non-integer mass dimension. 
One of the best static experimental methods to
determine the mass dimension $D$ of fractal materials 
is the box-counting method 
(see, for example \cite{BCM-1} and references therein).
It involves the selection of a box of size $R$ 
and counting the mass inside to estimate $D$ from corresponding 
power law relation $M \sim R^{D}$. 

Let us now consider a region $W$ of a fractal material
in the Euclidean space $\mathbb{R}^3$, with its
boundary denoted by $\partial W$.
Suppose that the medium in the region $W$ has 
a mass dimension $D$, and the medium on 
the boundary $\partial W$ has a dimension $d$. 
In general, the dimension $d$ is not equal to $(D-1)$ and 
it is not equal to $2$. 
The mass of the region $W$ in the fractal medium 
is denoted by $M_D(W)$. 
The fractality means that the mass 
in any region $W \subset \mathbb{R}^3$ 
increases slower than the 3D volume of this region, i.e.,
according to the power law $M_D(W) \sim R^{D}$, 
where $R$ is the radius of the ball used to measure $D$. 

A fractal material is called homogeneous 
if the power law $M_D(W) \sim R^{D}$ does not depend on 
the translation of the region $W$. 
In other words, for any two regions $W_1$ and $W_2$ of 
the homogeneous fractal material
with equal volumes $V_D(W_1)=V_D(W_2)$, 
the corresponding masses are equal $M_D(W_1)=M_D(W_2)$. 
A wide class of fractal media satisfies 
the homogeneous property. 
Many porous materials, polymers, colloid aggregates, and aerogels 
can be considered as homogeneous fractal materials.
However, the fact that a material is porous or random
does not necessarily imply that this material is fractal.
To describe fractal materials by a fractal integral 
continuum model, the fractality and homogeneity properties 
are implemented as follows: 

\begin{itemize} 
\item 
Homogeneity:
The local density of a homogeneous fractal material 
can be described 
by the constant density $\rho({\bf r})=\rho_0=const$. 
This property means that 
if $\rho({\bf r})=const$ and $V(W_1)=V(W_2)$, then $M_D(W_1)=M_D(W_2)$. 

\item 
Fractality:
The mass of the ball region $W$ of a fractal homogeneous material  
obeys a power law relation $M \sim R^{D}$, 
where $0<D<3$, and $R$ is the radius of the ball. 
If $V_n(W_1)= \lambda^n V_n(W_2)$ and $\rho({\bf r},t)=const$,
then fractality implies that $M_D(W_1)=\lambda^D M_D(W_2)$. 
\end{itemize}

These two conditions cannot be satisfied if the mass of 
the medium is described by an integral of integer order. 
In this case the mass is expressed by 
the fractional-order integral equation
\be \label{1-MW3} 
M_D(W,t)=\int_W \rho({\bf r},t) d V_D , \quad 
dV_D= c_3(D,{\bf r}) dV_3 ,
\ee
where ${\bf r}$ is a dimensionless vector variable. 
As already noted, $\rho({\bf r},t)$ is a distribution function, 
and $c_3(D,{\bf r})$ is a density of states in 
the Euclidean space $\mathbb{R}^3$.
The order of the integral in Eq. (\ref{1-MW3}) is defined 
by the fractal mass dimension of the material.
The kernel of the fractional integral operator describes 
a density of permitted states $c_3(D,{\bf r})$ in space,
and its symmetry is defined by the symmetry of the material structure.
The particular form (Riesz, Riemann-Liouville, etc.) 
of the function $c_3(D,{\bf r})$ is defined by the 
properties of the fractal material at hand. 
Note that the final field equations that relate 
the various physical variables of the system 
have a form that is independent of the numerical factor 
in the function $c_3(D,{\bf r})$. 
However the dependence on ${\bf r}$ is important in 
these equations. 
In addition, we note that for $D=2$, 
we have the fractal mass distribution
in 3D Euclidean space $\mathbb{R}^3$.
In general, this case is not equivalent to the distribution
on a 2D surface.

\subsection{Moment of inertia for fractal materials}

A method for calculating the moment of inertia of fractal 
materials has been suggested in \cite{IJMPB2005-2}.
The moment of inertia has two forms, 
a scalar form $I(t)$, which is used when the axis of rotation is known, 
and a more general tensor form that does not require knowing the axis of rotation. The scalar moment of inertia (often called simply the "moment of inertia") of a rigid body with density 
$\rho^{\prime}({\bf r}^{\prime},t)$ 
with respect to a given axis is defined by the volume integral
\be \label{1-(1)}
I^{\prime}(t)=\int_W \rho^{\prime}({\bf r}^{\prime},t) \; 
{\bf r}^{\prime \,2 }_{\perp} \; dV^{\prime}_3,
\ee 
where $({\bf r}^{\prime})^2_{\perp}$ is the square of
the perpendicular distance from 
the axis of rotation, and 
$dV^{\prime}_3=dx^{\prime}_1 dx^{\prime}_2 dx^{\prime}_3$.  
If ${\bf r}^{\prime}=x^{\prime}_k {\bf e}_k $ 
denotes the position vector from the origin to
a point ($x^{\prime}_k$, $k=1,2,3$, 
are components of ${\bf r}^{\prime}$), then 
the tensor form of the moment of inertia is
\be \label{1-(3)}
I^{\prime}_{kl}(t)=\int_W \rho^{\prime}({\bf r}^{\prime},t) \; 
\Bigl( ({\bf r}^{\prime})^2 \delta_{kl}- 
x^{\prime}_k x^{\prime}_l \Bigr) \; dV^{\prime}_3 ,
\ee
where $\delta_{kl}$ is the Kronecker delta. 
We note that the SI units of $I^{\prime}_{kl}$ is $kg \cdot m^2$, 
i.e., $[I^{\prime}_{kl}]=kg \cdot m^2$.

To generalize Eqs. (\ref{1-(1)}) and (\ref{1-(3)})
for fractional media, we express these equations through 
dimensionless coordinates.
We thus introduce the dimensionless variables
$x_k = x^{\prime}_k / l_0 , \quad {\bf r}={\bf r}^{\prime}/l_0$, 
where $l_0$ is a characteristic length scale, and 
write the density as
$\rho ({\bf r},t) = l^{3}_0 \, \rho^{\prime}( {\bf r} \, l_0 ,t)$
so its SI units is $m$, i.e., $[\rho]=kg$.
We then define the following moments of inertia  
$I_{kl}(t)= l^{-2}_0 I^{\prime}_{kl}(t)$, 
$I(t)=l^{-2}_0 I^{\prime}(t)$
to finaly obtain the relations
\be \label{1-(3)-2}
I(t)=\int_W \rho({\bf r},t) \;  {\bf r}^{2}_{\perp} \; dV_3,
\quad
I_{kl}(t) =\int_W \rho({\bf r},t) \; 
\Bigl( {\bf r}^2 \delta_{kl}- x_k x_l \Bigr) \; dV_3 , 
\ee
where $dV_3 =dx_1 dx_2 dx_3$ for Cartesian coordinates, and 
the variables $x_k$, $k=1,2,3$ are now dimensionless.
We note that the SI units of $I_{kl}$ is $kg$, i.e., 
$[I_{kl}]=kg $.
This representation allows us to generalize 
Eq. (\ref{1-(3)-2}) to fractal materials in the form 
\be \label{1-(3F)}
I^{(D)}(t)=\int_W \rho ({\bf r},t) \; 
{\bf r}^2_{\perp} \; dV_D , \quad
I^{(D)}_{kl}(t)=\int_W \rho({\bf r},t) \; 
({\bf r}^2 \delta_{kl}-x_kx_l)\; dV_D,
\ee
where $dV_D = c_3(D,{\bf r}) dV_3$ with
$D$ denoting, as usual, the mass dimension of 
the fractal material.

\subsection{Equilibrium equations for fractal materials} 

Let us now derive the equilibrium equations for 
a fractal material with mass dimension $D$.
Consider a finite region $W$ in the fractal material,
supporting a volume force and a surface force.
Let the density of force ${\bf f}({\bf r},t)$ be a function of 
the dimensionless vector ${\bf r}$, and time $t$.
The volume or mass force ${\bf F}_M(W)$, i.e.
the force acting on a region $W$ of a fractal 
medium with dimension $D$, is defined by 
\be \label{1-FM} 
{\bf F}_M (W) = \int_W \, {\bf f}({\bf r},t) \, dV_D.  \ee
The surface force ${\bf F}_S(W)$, i.e. the force acting 
on the surface $\partial W$ with dimension $d$, is defined by 
\be \label{1-FS} 
{\bf F}_S(W)=\int_{\partial W} \, {\bf \sigma}^n({\bf r},t) \, dA_{d},  \ee
where ${\bf \sigma}={\bf \sigma}({\bf r},t)$ is 
the traction vector on a surface with unit normal ${\bf n}$. 
As already mentioned, in general the dimension $d$ is not equal to $(D-1)$ and it is not equal to $2$. 
The resultant force that acts on the region $W$ is then
\be \label{1-ML1} 
{\bf F}_{\Sigma}(W) = {\bf F}_M(W) +{\bf F}_S(W) , \ee
and by substituting Eqs. (\ref{1-FM}) - (\ref{1-FS}) 
into Eq. (\ref{1-ML1}), we obtain
\be \label{1-ML2} 
{\bf F}_{\Sigma}(W) = 
\int_W \, {\bf f}({\bf r},t) \, dV_D +
\int_{\partial W} \, {\bf \sigma}^n({\bf r},t) \, dA_{d} . \ee 
This fractional integral equation represents the 
resultant force acting on any region $W$ of the fractal material. 
For $D=3$ and $d=2$, Eq. (\ref{1-ML2}) gives 
the usual equation for the resultant force in a 
non-fractal continuum. 
The force equilibrium condition for the region $W$ 
requires ${\bf F}_{\Sigma}(W) =0$. 
Therefore, we have the fractional integral equation of equilibrium 
\be \label{EE-FM-1} 
\int_W \, {\bf f}({\bf r},t) \, dV_D+
\int_{\partial W} \, {\bf \sigma}^n({\bf r},t) \, dA_{d} =0 .
\ee
In component form, this equation reads
\be \label{EE-FM-1b} 
\int_W \, f_k({\bf r},t) \, dV_D+
\int_{\partial W} \, \sigma^n_k({\bf r},t) \, dA_{d} =0 ,
\ee
where we use 
${\bf f}= f_k {\bf e}_k$ and ${\bf \sigma}^n=  \sigma^n_{k} {\bf e}_k$.
Using the normal vector ${\bf n}=n_j {\bf e}_j$, 
we can represent $\sigma^n_{k}$ in the form
$\sigma^n_{i} = \sigma_{ij} n_j$, where $\sigma_{ij}$ is the stress tensor.


The differential form of equilibrium equations 
follows directly from Eq. (\ref{EE-FM-1b}). 
Using the generalization of the Gauss theorem for fractal media \cite{AP2005-2}, the surface integral can be represented as 
\be \label{Gauss}
\int_{\partial W} {\bf \sigma}^n \, dA_{d} =
\int_{\partial W} c_2(d,{\bf r}) \, {\bf \sigma}^n \, dA_{2} 
= \int_W  \frac{\partial 
(c_2(d,{\bf r}) \, {\bf \sigma}_l)}{\partial x_l} c^{-1}_3(D,{\bf r}) \, dV_D=
\int_W  \nabla^{(D,d)}_l {\bf \sigma}_l \, dV_D , 
\ee
where a generalization of the nabla operator for fractal materials \cite{AP2008} was also used in the form 
\be \label{1-nablaD}
\nabla^{(D,d)}_k B = c^{-1}_3(D,{\bf r}) \frac{\partial (c_2(d,{\bf r}) B)}{\partial x_k} ,
\ee
where $B=B({\bf r})$ is a function of the coordinates. 
This operator will be called "fractal-nabla" operator. 
We note that the operator given by Eq. (\ref{1-nablaD}) 
is not a fractional derivative \cite{KST} or 
an operator on a fractal set \cite{Kugami}.
For example, if we use the density of states $c_3(D,{\bf r})$ and $c_2(d,{\bf r})$ in the form
\be \label{c3R}
c_3(D,{\bf r})=\frac{2^{3-D} \Gamma(3/2)}{\Gamma(D/2)} |{\bf r}|^{D-3} ,
\ee
\be \label{c2R} 
c_2(d,{\bf r}) = \frac{2^{2-d}}{\Gamma(d/2)} |{\bf r}|^{d-2} , 
\ee
then the "fractal-nabla" operator is given by
\be \label{1-nablaD2}
\nabla^{(D,d)}_k B =
\frac{2^{D-d-1} \Gamma(D/2)}{\Gamma(3/2) \Gamma(d/2)} \,
|{\bf r}|^{3-D} \frac{\partial }{\partial x_k} 
\left( |{\bf r}|^{d-2} B \right). \ee
For non-fractal materials ($D=3$ and $d=2$), we have
\[ \nabla^{(3,2)}_k B= \frac{\partial B}{\partial x_k} . \]
We note that the rule of term-by-term differentiation 
for the operator $\nabla^{(D,d)}_k$ is not satisfied, i.e.
\[ \nabla^{(D,d)}_k(BC)\not= B\nabla^{(D,d)}_k(C)+C\nabla^{(D,d)}_k(B) . \]
The operator $\nabla^{(D,d)}_k$ satisfies the following rule
\be \label{1-7} 
\nabla^{(D,d)}_k (BC)=B\nabla^{(D,d)}_k (C)+c(D,d,{\bf r}) \, C \, \nabla^1_k B , \ee 
where 
\[ c(D,d,{\bf r}) =c^{-1}_3(D,{\bf r}) c_2(d,{\bf r}) . \]
For example, the density of states given by 
Eqs. (\ref{c3R}) and (\ref{c2R}), can be expressed as
\[ c(D,d,{\bf r}) =
\frac{2^{D-d-1} \Gamma(D/2)}{\Gamma(3/2) \Gamma(d/2)} \, |{\bf r}|^{d+1-D} . \]
Note that, in general, $\nabla^{(D,d)}_k (1)\not=0$ since
\[ \nabla^{(D,d)}_k (1)=c(D,d,{\bf r}) \, (d-2) \frac{x_k}{r^2} . \]

Using now Eq. (\ref{Gauss}), Eq. (\ref{EE-FM-1}) takes the form
\be \label{EE-FM-2} 
\int_W  \left( {\bf f} + \nabla^{(D,d)}_l {\bf \sigma}_l \right) \, dV_D =0
, \ee 
or in components form (with ${\bf f}=f_k {\bf e}_k$, and 
${\bf \sigma}^n_l=\sigma_{kl} {\bf e}_k$), we have
\be \label{EE-FM-3}
\int_W \Bigl( f_k + \nabla^{(D,d)}_l \sigma_{kl} \Bigr) \, dV_D =0 , 
\quad (k=1,2,3) . \ee
This equation is satisfied for all regions $W$. 
As a result, we have 
\be \label{EE-FM-4}
\nabla^{(D,d)}_l \sigma_{kl} +f_k =0 , \quad (k=1,2,3) . 
\ee
Using the usual notation, we have
\be \label{EE-FM-5}
c^{-1}_3(D,{\bf r}) \, D^1_{x_l} \, \Bigl( c_2(d,{\bf r}) \, \sigma_{kl} \Bigr) +f_k =0 , \quad (k=1,2,3) . 
\ee
These are the differential equations of equilibrium 
for fractal materials. 


Let us derive next, the equilibrium equation for 
the moment of forces. 
The moment ${\bf M}_M(W)$ of the mass force (\ref{1-FM}), 
can be written as 
\be \label{1-MM} 
{\bf M}_M (W) = \int_W \, [{\bf r},{\bf f}] \, dV_D . \ee
The moment ${\bf M}_S(W)$ of the surface force (\ref{1-FS}) 
is given by
\be \label{1-MS} 
{\bf M}_S(W)=\int_{\partial W} \, [{\bf r},{\bf \sigma}^n] \, dA_{d}.  \ee
In Eqs. (\ref{1-MM}) and (\ref{1-MS}), 
the brackets $ [ \ . \ , \ . \ ]$ denotes
vector product of vector fields.
The resultant moment ${\bf M}_{\Sigma}(W)$ is the sum
\be \label{M-Sum-1} 
{\bf M}_{\Sigma}(W) = {\bf M}_M(W) +{\bf M}_S(W) . \ee
Substituting Eqs. (\ref{1-MM}) - (\ref{1-MS}) 
into Eq. (\ref{M-Sum-1}), we obtain
\be \label{M-Sum-2} 
{\bf M}_{\Sigma}(W) = 
\int_W \, [{\bf r},{\bf f}] \, dV_D +
\int_{\partial W} \, [{\bf r},{\bf \sigma}^n] \, dA_{d} . \ee 
The equilibrium condition for 
the region $W$ surrounded by its surface $\partial W$
of a fractal material leads to ${\bf M}_{\Sigma}(W) =0$,
yielding the fractional integral equation  
\be \label{M-EQ-1} 
\int_W \, [{\bf r},{\bf f}] \, dV_D +
\int_{\partial W} \, [{\bf r},{\bf \sigma}^n] \, dA_{d}=0 .
\ee
In component form, this equation reads
\be \label{M-EQ-2} 
\int_W \, \epsilon_{ijk} \, x_j \,f_k \, dV_D +
\int_{\partial W} \, \epsilon_{ijk} \, x_j \, \sigma_{kl} \, n_l \, dA_{d} ,
\ee
where $\epsilon_{ijk}$ is the Levi-Civita symbol.
Using then the generalization of Gauss theorem for fractal materials given by Eq. (\ref{Gauss}), we obtain
\[
\int_{\partial W} \, \epsilon_{ijk} \, x_j \, \sigma_{kl} \, n_l \, dA_{d} =
\int_{\partial W} \, \epsilon_{ijk} \, x_j \, \sigma_{kl} \, n_l \, c_2(d, {\bf r})\, dA_2 =
\]
\[ 
= \int_{\partial W} \, \epsilon_{ijk} \, D^1_l \Bigl(
 x_j \, \sigma_{kl} \, c_2(d, {\bf r}) \Bigr) \, dV_3
= \int_{\partial W} \, \epsilon_{ijk} \, c^{-1}_3(D, {\bf r}) \, D^1_l \Bigl(
 x_j \, c_2(d, {\bf r}) \, \sigma_{kl}  \Bigr) \, dV_D =
\]
\[
= \int_{\partial W} \, c(D,d,{\bf r}) \epsilon_{ilk} \,  \, \sigma_{kl} \, dV_D +
\int_{\partial W} \, \epsilon_{ijk} \, x_j \, \nabla^{(D,d)}_l \, \sigma_{kl} \, dV_D =
\]
\be \label{final}
= \int_{\partial W} \, c(D,d,{\bf r}) \epsilon_{ilk} \,  \, \sigma_{kl} \, dV_D -
\int_{\partial W} \, \epsilon_{ijk} \, x_j \, f_k \, dV_D ,
\ee
where equation (\ref{EE-FM-4}) is also used.
Substitution of Eq. (\ref{final}) into 
Eq, (\ref{M-EQ-2}) gives
\be
\int_{\partial W} \, c(D,d,{\bf r}) \epsilon_{ilk} \,  \, \sigma_{kl} \, dV_D =0 .
\ee
This equation is satisfied for all regions $W$. 
Therefore we have the condition 
\be
\epsilon_{ijk} \sigma_{kj}=0 , 
\ee 
or, equivalent, 
\be
\sigma_{ij}=\sigma_{ji} .
\ee
This equilibrium equation for the moment of the force in fractal materials is the same as for the non-fractal case,
and suggests that the stress tensor is symmetric.

\subsection{Conservation laws for fractal materials}

In the framework of fractional integral continuum model, 
the fractional conservation laws 
for fractal media have been derived in \cite{AP2005-2} 
(see also \cite{MOS-3,TarasovSpringer}). 
For future reference, the differential equations of 
the conservation laws are also summarized below:

(1) The conservation law for mass
\be \label{1-1eq} 
\left(\frac{d}{dt}\right)_{(D,d)} \rho 
=-\rho \, \nabla^{(D,d)}_k u_k . \ee

(2)
The conservation law for momentum 
\be \label{1-2eq} 
\rho \, \left(\frac{d}{dt}\right)_{(D,d)} u_k=
f_k+ \nabla^{(D,d)}_l \sigma_{kl} . \ee

(3) The conservation law for energy 
\be \label{1-3eq} 
\rho \, \left(\frac{d}{dt}\right)_{(D,d)} e = 
c(D,d,{\bf r}) \, 
\sigma_{kl} \, D^1_l u_k + \nabla^{(D,d)}_k q_k . \ee

It is noted that these equations are differential equations 
with derivatives of integer order (see Eq. (\ref{1-nablaD2})).
It is also pointed out that 
the generalized total time derivative is defined by
\be 
\left(\frac{d}{dt}\right)_{(D,d)}= \frac{\partial}{\partial t}+
c(D,d,{\bf r}) \, u_l \, D^1_l , \ee
where $r=|{\bf r}|$, $x_k$, $k=1,2,3$, are dimensionless variables, the operator $D^1_l$ is defined 
as usual by $D^1_l=\partial / \partial x_k$, and
\[ c(D,d,{\bf r}) = c^{-1}_3(D,{\bf r}) \, c_2(d,{\bf r}) . \]
The above listed differential equations of balance 
for the density of mass, the density of momentum, and 
the density of internal energy 
make up a set of five equations, which are not closed.
In addition to the fields
$\rho({\bf r},t)$, $u({\bf r},t)$, $e({\bf r},t)$, 
equations (\ref{1-2eq}) and (\ref{1-3eq}) include 
the tensor of stress $\sigma_{kl}({\bf r},t)=\sigma_{lk}({\bf r},t)$
and the vector of thermal flux $q_k({\bf r},t)$. 
It is also remarked that the conservation laws 
for fractal media, which are suggested in \cite{MOS-3} 
are different from the conservation laws given by Eqs.
(\ref{1-1eq}-\ref{1-3eq}) 
derived in \cite{AP2005-2,TarasovSpringer}.
In \cite{MOS-3} all equations contain the derivatives 
$c^{-1}_1(\alpha_{x_i},x_i) \, D^{1}_{x_i}$ only,
where the density of states
$c^{-1}_1(\alpha_{x_i},x_i)$
can be considered as
$c^{-1}_3(D,{\bf r}) c_2(D-\alpha_{x_i},{\bf r}- x_i {\bf e}_i)$.
Equations (\ref{1-1eq}-\ref{1-3eq}) contain two types of derivatives: $D^1_l$ and $\nabla^{(D,d)}_k$.

\subsection{Constitutive relations for fractal materials}

For the theory of non-fractal gradient elasticity of 
isotropic materials the constitutive relations  
\cite{A1992}-\cite{RA1993} has the form
\be \label{GCR-1}
\sigma_{ij} = \Bigl( \lambda  \varepsilon_{kk} \delta_{ij} + 2 \mu \varepsilon_{ij} \Bigr)
- l^2 \, \Delta \, \Bigl( \lambda  \varepsilon_{kk} \delta_{ij} + 2 \mu \varepsilon_{ij} \Bigr) ,
\ee
where $\sigma_{ij}$ and $\varepsilon_{ij}$ are the stress and
strain tensors and $l$ denotes an internal length.
As usual, $\lambda$ and $\mu$ are the Lame coefficients; and
$\Delta$ is the Laplace operator defined by 
the scalar product of the nabla operators
\be
\Delta = (\nabla,\nabla) =\sum^2_{k} (\nabla_k)^2 .
\ee
It is easy to see that the balance equations 
for fractal media considered herein contain
in addition to the usual derivatives $D^1_k$
the "fractal-nabla" operator $\nabla^{(D,d)}_k $ 
of Eq. (\ref{1-nablaD}),
\be \label{nabla-D}
\nabla^{(D,d)}_k \Bigl( \ . \ \Bigr)
= c^{-1}_3(D,{\bf r}) \nabla_k \Bigl( c_2(d,{\bf r})  \ . \ \Bigr) 
\ee
that takes into account the density of states of fractal media with non-integer mass dimensions.
Therefore, we can assume that corresponding generalizations 
of constitutive relations can be obtained 
by the replacement of the usual nabla operator by 
the "fractal-nabla" operator. 
For example, a fractal generalization of
the gradient elasticity model given by Eq.
(\ref{GCR-1}) can be represented by the constitutive relations
in the form
\be \label{GCR-1F}
\sigma_{ij} = \Bigl( \lambda  \varepsilon_{kk} \delta_{ij} + 2 \mu \varepsilon_{ij} \Bigr)
- l^2_F \, \Delta^{(D,d)} \, \Bigl( \lambda  \varepsilon_{kk} \delta_{ij} + 2 \mu \varepsilon_{ij} \Bigr) ,
\ee
where we use the "fractal-Laplacian" that is defined by
\be
\Delta^{(D,d)} = \Bigl(\nabla^{(D,d)},\nabla^{(D,d)} \Bigr) =\sum^2_{k} \Bigl(\nabla^{(D,d)}_k \Bigr)^2 .
\ee
For non-fractal materials, we have $D=3$, $d=2$ and
$\Delta^{(3,2)}=\Delta$.
More generally, we can assume that
the constitutive relations for fractal materials are
of the form
\be \label{GCR-2F}
\sigma_{ij} = \Bigl( \lambda  \varepsilon_{kk} \delta_{ij} + 2 \mu \varepsilon_{ij} \Bigr)
- l^2_S \, \Delta \, \Bigl( \lambda  \varepsilon_{kk} \delta_{ij} + 2 \mu \varepsilon_{ij} \Bigr)
- l^2_F \, \Delta^{(D,d)} \, \Bigl( \lambda  \varepsilon_{kk} \delta_{ij} + 2 \mu \varepsilon_{ij} \Bigr) ,
\ee
where two types of Laplacians are taken into account.

In general, fractal materials cannot be defined 
as media distributed over a fractal set.
Naturally, in real materials the fractal structure 
cannot be observed on all scales. 
Materials demonstrate fractality only in a range of scales  
$R_{min} < R < R_{max}$. 
If the sample material has a size $R_{S}$ greater than $R_{max}$, or the region of scales  $[R_{min},R_{max}]$ is narrow, 
then the material is "semi-fractal" material.
The parameter $l^2_S$ in constitutive relation 
given by Eq. (\ref{GCR-2F})
is a measure of spatial non-fractality of the material,
whereas the parameter $l^2_F$ is a measure of spatial 
fractality of material for
the fractal gradient elasticity theory considered herein.
Which of the two models of Eq. (\ref{GCR-1F}) or 
Eq. (\ref{GCR-2F}) is more appropriate to describe 
a particular fractal material, must be determined experimentally.

\subsection{Strain-displacement relation for fractal materials}

In \cite{MOS-3}-\cite{MOS-5} it is postulateed that  
the strain $\varepsilon_{ij}$ for small deformations 
of fractal materials 
is given in terms of the displacement $u_k$ by the equation
\be \label{Def-MOS-1}
\varepsilon_{ij} =\frac{1}{2}  \Bigl( c^{-1}_1(\alpha_{x_i},x_i) \, D^{1}_{x_i} u_j +
c^{-1}_1(\alpha_{x_j},x_j) \, D^{1}_{x_j} u_i \Bigr) .
\ee
The one-dimensional analogue of Eq. (\ref{Def-MOS-1})
has been considered in \cite{MOS-4,MOS-5} in the form
\be \label{Def-MOS-2}
\varepsilon (x)=  c^{-1}_1(\alpha,x) \, D^{1}_{x} \, u(x) ,
\ee
where $c_1(\alpha,x)$ is the density of states.
As a basis for using this definition, 
reference is made the differential form of a linear element
$dl_{\alpha} = c^{-1}_1(\alpha_x,x) dx$, which
takes into account the 1D density of states. 
Another argument \cite{MOS-4,MOS-5} to support this choise  
is a possibility to obtain the same 1D
elastic wave equation from a variational principle, as 
the wave equation obtained from the balance equations.
However, it is not quite clear the necessity to consider  
the density of states in the definition of the strain.

It thus seems that the definitions given by Eqs. 
(\ref{Def-MOS-1}) or (\ref{Def-MOS-2}) are not  
sufficiently rigorously justified.
The inclusion of the density of states 
$c_1(\alpha,x)$ into the strain-displacement relation
looks like an artificial reception.
The relation between the strain tensor 
$\varepsilon_{ij}$ and the displacement vector $u_k$ 
should be derived directly from the relevant distance changes 
(for example, see Section 1.1 of \cite{LL}),
and this relation should not be postulated in definition.
For example, in the 1D case, the strain-displacement relation 
for fractal materials should be derived from the equation
\be \label{Dist}
(dl^{\prime}_{\alpha})^2 = (dl_{\alpha})^2 \, (1+ 2\varepsilon (x))  
\ee 
that describes the deformation of a linear element 
$dl_{\alpha} = c_1(\alpha_x,x) dx$ of 1D fractal medium. 
From Eq. (\ref{Dist}) it is apparent that the strain $\varepsilon(x)$ 
does not contain the density of states $c_1(\alpha,x)$.
The relation between strain and displacement should 
define the deformation of a volume element
$dV_D = c_3(D,{\bf r}) \,dV_3$ of a fractal material 
through the condition
\be
dV^{\prime}_D = dV_D [1+ \varepsilon_{11}(x) + \varepsilon_{22}(x)+\varepsilon_{33}(x) ] ,
\ee 
which is the fractal analogue of Eq. (1.6) of \cite{LL},
we see that $\varepsilon_{ii}(x)$ 
does not contain the density of states also.

\subsection{Variational principle for fractal materials}

Another way to derive the governing equations for
fractional integral continuum models for fractal materials 
is the use of variational principles. 
A holonomic variational principle for fractal materials 
has been suggested in \cite{MPLB2005-1,Physica2005}
in the framework of a fractional integral continuum model.
Variational principles for fractal elasticity  
are also considered in \cite{MOS-1,MOS-2}.
The equation for fractal elasticity 
can be derived as the Euler-Lagrange equations 
from a holonomic functional.

Let us consider a fractional integral continuum model 
for fractal materials in $\mathbb{R}^3$ that 
is described by the action 
\be \label{action-1F0}
S_F[u] = \int dt \, \int_{\mathbb{R}^3} dV_D \, 
\mathcal{L}(u_i, u_{i,t}, u_{i,k} , u_{i,kl}, u_{i,klm}) 
\ee
with Lagrangian $\mathcal{L}(u_i, u_{i,t}, u_{i,k} , u_{i,kl}, u_{i,klm} )$, where $u_i=u_i({\bf r},t)$ is the displacement vector.  
To take into account the fractality of the material 
in coordinate space $\mathbb{R}^3$, we use
\[ dV_D = c_3(D,{\bf r}) dV_3 , \]
where the function $c_3(D,{\bf r})$ describes the density of states in $\mathbb{R}^3$.
Note that $x$, $y$, $z$ and ${\bf r}$ are dimensionless variables.

The variation of the action functional given by Eq. (\ref{action-1F0}) is
\[ \delta S_F[u] = 
\int dt \, \int_{\mathbb{R}^3} dV_D \,  
 \delta \mathcal{L} = \int dt \, \int_{\mathbb{R}^3} dV_D \, 
\Bigl[ \frac{\partial\mathcal{L}}{\partial u_i} \delta u_i + 
 \left(\frac{\partial \mathcal{L}}{\partial u_{i,t}}\right) \, \delta u_{i,t} + \]
\be \label{VarS-1F0}
+\left(\frac{\partial \mathcal{L}}{\partial u_{i,k}}\right) \, \delta u_{i,k} 
+\left(\frac{\partial \mathcal{L}}{\partial u_{i,kl}}\right) \, \delta u_{i,kl}  
+  \left(\frac{\partial \mathcal{L}}{\partial u_{i,klm}}\right) \, \delta u_{i,klm})  \Bigr] .
\ee
If the fractal material is not subjected to 
non-holonomic constraints, then 
the variation and fractional derivatives commute, 
\[
\delta u_{i,t} = D^{1}_t (\delta w) , \quad
\delta u_{i,k} = \nabla_k (\delta w) , \quad
\delta u_{i,kl}  = \nabla_k \nabla_l (\delta w) , \quad
\delta u_{i,klm}  = \nabla_k \nabla_l \nabla_m (\delta w)  .
\]
Using integration by parts, 
we can express Eq. (\ref{VarS-1F0}) in the form
\[ \delta S_F [u] = \int dt \, \int_{\mathbb{R}^3} dV_3 \,  
\Bigl[ c_3(D,{\bf r}) \, \frac{\partial\mathcal{L}}{\partial u_{i}} \delta u_{i} - 
D^{1}_t \, \left( c_3(D,{\bf r}) \, \frac{\partial \mathcal{L}}{\partial u_{i,t}} \right) + \]
\be \label{VarS-2F0}
- \nabla_k \, \left( c_3(D,{\bf r}) \, 
 \frac{\partial \mathcal{L}}{\partial u_{i,k}}\right)   
+ \nabla_k \nabla_l \, \left( c_3(D,{\bf r}) \, 
 \frac{\partial \mathcal{L}}{\partial u_{i,kl}}\right)   
- \nabla_k \nabla_l \nabla_m \, \left( c_3(D,{\bf r})  \, \frac{\partial \mathcal{L}}{\partial u_{i,klm} }\right) \Bigr] \, \delta u_{i} .
\ee
Then, the stationary action principle, in the form of the holonomic variational equation $\delta S_F[u] =0$, 
gives the Euler-Lagrange equations for the fractional 
integral continuum model of the fractal material 
considered in the form
\[
\frac{\partial\mathcal{L}}{\partial u_{i}} \delta u_{i} - 
D^{1}_t \, \left( \frac{\partial \mathcal{L}}{\partial u_{i,t}}\right) 
- c^{-1}_3(D,{\bf r}) \, \nabla_k \, \left( c_3(D,{\bf r}) \, 
 \frac{\partial \mathcal{L}}{\partial u_{i,k}} \right)+
 \]
\be \label{ELE-1F0}
+ c^{-1}_3(D,{\bf r}) \, \nabla_k \nabla_l \, \left( c_3(D,{\bf r}) \, 
 \frac{\partial \mathcal{L}}{\partial u_{i,kl}} \right)   
- c^{-1}_3(D,{\bf r}) \, \nabla_k \nabla_l \nabla_m \, \left( c_3(D,{\bf r})  \, \frac{\partial \mathcal{L}}{\partial u_{i,klm} }\right) =0 .
\ee
It follows that a mathematical model for a fractal material 
is entirely determined by the choice of the Lagrangian.
We demonstrate an application of this approach 
by considering the example of the Euler-Bernoulli fractal beam 
in the next section.


\newpage
\section{Gradient elasticity model for fractal beam}


In this section we derive a gradient elasticity model 
for fractal materials in the form of 
the Euler-Bernoulli beam equation of motion
by using the holonomic variational principle for fractal media \cite{MPLB2005-1,Physica2005}.
We will consider the gradient fractal beam by using 
the fractional integral continuum approach suggested
in \cite{AP2005-2,PLA2005-1,TarasovSpringer}. 
In this connection, it is noted that 
a non-gradient fractal beam has been considered 
in \cite{MOS-4,MOS-5} in the framework of 
a fractional integral continuum model.

\subsection{Variational equation for 1-dimesional model of fractal materials}

Let us consider a 1D fractional continuum model 
for fractal materials described by the action 
\be \label{action-1F}
S_F [w] = \int dt \, \int dl_{\alpha_x} \, 
\mathcal{L}(x, t, w, D^{1}_t w, D^{2}_x w , D^{3}_x w) 
\ee
with Lagrangian $\mathcal{L}(x, t, w, D^{1}_t w, 
D^{2}_x w , D^{3}_x w) $, where 
$dl_{\alpha_x} = c_1(\alpha_x,x) \, dx$
and $x$ is dimensionless.
The function $c_1(\alpha_x,x)$ denotes the density of 
states along the $x$-axis.
For the Euler-Bernoulli fractal beam model, 
the field $w(x)=u_y(x)$ is the curve that describes the deflection of the beam in the $y$ direction at some position $x$.

The variation of the action functional 
given by Eq. (\ref{action-1F}) is
\[ \delta S_F [w] = 
\int dt \, \int dx \, c_1(\alpha_x,x) \, 
 \delta \mathcal{L} = \int dt \, \int dx \, c_1(\alpha_x,x) \, 
\Bigl[ \frac{\partial\mathcal{L}}{\partial w} \delta w + 
 \left(\frac{\partial \mathcal{L}}{\partial D^{1}_t w}\right) \, \delta (D^{1}_t w) + \]
\be \label{VarS-1F}
+\left(\frac{\partial \mathcal{L}}{\partial D^{2}_x w}\right) \, \delta (D^{2}_x w)  
+  \left(\frac{\partial \mathcal{L}}{\partial D^{3}_x w}\right) \, \delta (D^{3}_x w)  \Bigr] .
\ee
If non-holonomic constraints are not involved,  
the variation and fractional derivatives commute, i.e.
\[
\delta (D^{1}_t w)  = D^{1}_t (\delta w) , \quad
\delta (D^{2}_x w)  = D^{2}_x (\delta w) , \quad
\delta (D^{3}_x w)  = D^{3}_x (\delta w)  .
\]
Using integration by parts, we 
express Eq. (\ref{VarS-1F}) in the form
\[ \delta S_F [w] = \int dt \, \int dx \, 
\Bigl[ c_1(\alpha_x,x) \, \frac{\partial\mathcal{L}}{\partial w} \delta w - 
D^{1}_t \, \left( c_1(\alpha_x,x) \, \frac{\partial \mathcal{L}}{\partial D^{1}_t w}\right) + \]
\be \label{VarS-2F}
+D^{2}_x  \, \left(c_1(\alpha_x,x) \, 
 \frac{\partial \mathcal{L}}{\partial D^{2}_x w}\right)   
- D^{3}_x \, \left( c_1(\alpha_x,x) \, \frac{\partial \mathcal{L}}{\partial D^{3}_x w}\right) \Bigr] \, \delta w  .
\ee
The stationary action principle implies
the holonomic variational equation $\delta S_F[w] =0$.
This equation gives the Euler-Lagrange equation in the form
\be \label{ELE-1F}
\frac{\partial\mathcal{L}}{\partial w} - 
c_1(\alpha_x,x) \, D^{1}_t \, \left(\frac{\partial \mathcal{L}}{\partial D^{1}_t w}\right)  
+ D^{2}_x \, \left( c_1(\alpha_x,x) \,\frac{\partial \mathcal{L}}{\partial D^{2}_x w}\right) 
- D^{3}_x \, \left( c_1(\alpha_x,x) \,\frac{\partial \mathcal{L}}{\partial D^{3}_x w}\right) = 0 .
\ee
This equation describes the fractional continuum model of 
a fractal material distributed in $\mathbb{R}^1$ 
with dimension $\alpha_z$.


\subsection{Euler-Lagrange equation for 
the Euler-Bernoulli fractal beam}

The Lagrangian for the Euler-Bernoulli fractal beams 
has the form 
\[ \mathcal{L}(x, t, w, D^{1}_t w, 
D^{2}_x w , D^{3}_x w) = \frac{1}{2} \mu 
\left( D^{1}_t w(x,t) \right)^2 +
\frac{1}{2} (E\, I^{(d)}) \, \left( D^{2}_x w(x,t) \right)^2 -\]
\be \label{Lag-1F}
- \frac{1}{2} (E\, I^{(d)}) \, l^2_F(d)  
\left( D^{3}_x w(x,t) \right)^2  - q(x,t) w(x,t) .
\ee
The first term represents the kinetic energy, 
where $\mu = \rho \, A$ is the mass per unit length; 
the second one represents the potential energy due to 
an internal forces (when considered with a negative sign); 
and the third term represents the potential energy due 
to the external load $q(x,t)$.
Note that ($x$, $y$, $z$) are dimensionless variables, 
and $l^2_F(d)$ is a dimensionless parameter. 

The Lagrangian looks similar the usual Lagrangian 
for an Euler-Bernoulli gradient elastic beam. 
A difference is in the presence of the moment of inertia 
$I^{(d)}$ of the fractal material only.
In the Lagrangian we used the second moment 
of area ($I^{(d)}=I^{(d)}_z$) of the fractal beam's cross-section 
defined by
\be 
I^{(d)}=I^{(d)}_z = \iint_A y^2 \; dA_{x}(d) ,
\ee
where we take into account the density of states $c_2(d,y,z)$ 
in the expression of a fractal surface differential element, i.e.
\[ dA_{x}(d) = c_2(d,y,z) \, d A_x . \]


In \cite{MOS-4,MOS-5} it has been suggested 
to use the derivatives $c^{-1}_1(\alpha_x,x) \, D^{1}_x $ 
instead of the usual derivatives $D^1_x$ for fractal materials.
If we use the derivatives $c^{-1}_1(\alpha_x,x) \, D^{1}_x $ 
instead of $D^1_x$ for fractal materials 
according to \cite{MOS-4,MOS-5}, then
the Lagrangian for Euler-Bernoulli fractal beams 
takes the following form
\[ \mathcal{L}(x, t, w, D^{1}_t w, 
D^{2}_x w , D^{3}_x w) = \frac{1}{2} \mu 
\left( D^{1}_t w(x,t) \right)^2 +
\frac{1}{2} (E\, I^{(d)}) \, \left( (c^{-1}_1(\alpha_x,x) \, D^{1}_x )^2 w(x,t) \right)^2 -\]
\be \label{Lag-2F}
- \frac{1}{2} (E\, I^{(d)}) \, l^2_F(d)  
\left( (c^{-1}_1(\alpha_x,x) \, D^{1}_x)^3 w(x,t) \right)^2  - q(x,t) w(x,t) .
\ee


Using the Lagrangian (\ref{Lag-1F}), the corresponding terms 
in the relevant Euler-Lagrange equation,
i.e. Eq. (\ref{ELE-1F}), are
\be \label{Sub-1}
\frac{\partial\mathcal{L}}{\partial w} = - q(x,t)  \qquad
\frac{\partial \mathcal{L}}{\partial D^{1}_t w(x,t) } =  
\mu \, D^{1}_t w(x,t) , 
\ee
\be \label{Sub-2}
\frac{\partial \mathcal{L}}{\partial D^{2}_x w(x,t)} = (E\, I^{(d)})\, D^{2}_x w(x,t)  , \qquad
\frac{\partial \mathcal{L}}{\partial D^{3}_x w(x,t)} = (E\, I^{(d)}) \, l^2_F(d) \, D^{3}_x w(x,t) .
\ee
Substitution of Eqs. (\ref{Sub-1}) and (\ref{Sub-2}) 
into Eq. (\ref{ELE-1F}) gives
\[ 
\mu \, D^{2}_t w +
c^{-1}_1(\alpha_x,x) \,D^{2}_x \, \Bigl( c_1(\alpha_x,x) \, (E\, I^{(d)}) \, (D^{2}_x) w \Bigr) - \]
\be \label{FEB-1F}
- l^2_F (d) \, c^{-1}_1(\alpha_x,x) \, D^{3}_x \, \Bigl( c_1(\alpha_x,x) \, (E\, I^{(d)}) \,  D^{3}_x w \Bigr) 
- q(x,t) = 0 ,
\ee
which is the governing equation of motion for 
a fractal Euler-Bernoulli beam.
For a non-fractal beam, we have $\alpha_x=1$, 
$c^{-1}_1(\alpha_x,x)=1$, and the standard 
gradient elasticity Euler-Bernoulli beam equation is recovered
\be \label{FEB-1NF}
\mu \, D^{2}_t w +
D^{2}_x \, \Bigl( (E\, I) \, (D^{2}_x) w \Bigr) 
- l^2_s \, D^{3}_x \, \Bigl( (E\, I) \,  D^{3}_x w \Bigr) 
- q(x,t) = 0 ,
\ee
where the beam can be non-homogeneous, and $E$ and $I$ may depend on $x$.

If the fractal beam is homogeneous (see Section 4.2), then $E$ and $I^{(d)}$ are independent of $x$, and the beam equation 
has a simpler form
\[
\mu \, D^{2}_t w + (E\, I^{(d)}) \,
c^{-1}_1(\alpha_x,x) \,D^{2}_x \, \Bigl( c_1(\alpha_x,x) \, D^{2}_x w \Bigr) - \]
\be \label{FEB-2F}
- l^2_F (d) \, (E\, I^{(d)}) \, c^{-1}_1(\alpha_x,x) \, D^{3}_x \, \Bigl( c_1(\alpha_x,x) \, D^{3}_x w \Bigr) 
- q(x,t) = 0 .
\ee
This equation can be expressed as
\be \label{FEB-3F}
\mu \, D^{2}_t w +
E \, I^{(d)} \, \mathbb{D}^{4}_{x,\alpha_x}  w  -
l^2_F (d) \, E \, I^{(d)} \, \mathbb{D}^{6}_{x,\alpha_x}  w 
- q(x,t) = 0 ,
\ee
where we have used the notation
\be \label{not-1}
\mathbb{D}^{2n}_{x,\alpha_x} = c^{-1}_1(\alpha_x,x) \,D^{n}_x \,  c_1(\alpha_x,x) \, D^{n}_x .
\ee
If $\alpha_x=1$, then $ c_1(\alpha_x,x) =1$ and
$\mathbb{D}^{2n}_{x,\alpha_x} =D^{2n}_x$. 

Using the Lagrangian given by Eq. (\ref{Lag-2F}), 
the corresponding Euler-Lagrange equation 
has the form of Eq. (\ref{FEB-2F}), 
where the derivatives $\mathbb{D}^{2n}_{x,\alpha_x}$
are replaced by 
\[ \partial^{2n}_{x,\alpha_x} = 
(c^{-1}(\alpha_x,x) \, D^1_{x})^{2n} , \]
such that
\be \label{FEB-3F-MOS}
\mu \, D^{2}_t w +
E \, I^{(d)} \, \partial^{4}_{x,\alpha_x} w  -
l^2_F (d) \, E \, I^{(d)} \, \partial^{6}_{x,\alpha_x} w 
- q(x,t) = 0 .
\ee
For non-fractal materials, we have $\alpha_x=1$ and 
Eqs. (\ref{FEB-3F}), (\ref{FEB-3F-MOS}) have the form
\be \label{FEB-4F}
\mu \, D^2_t w + E\, I \, D^4_x w - 
E\, I \, l^2_F(2) \, D^6_x w - q(x,t) = 0 .
\ee
This is the gradient elasticity Euler-Bernoulli beam equation
for media without fractional non-locality, memory and fractality \cite{AA2011}.

\subsection{Second moment of area for fractal beam}

In this section, we give an example of computation 
a second moment of the fractal beam's cross-section
by the method suggested in \cite{IJMPB2005-2}.
Let us consider a homogeneous fractal beam 
with circular cross-section. 
The second moment $I^{(d)}=I^{(d)}_z$ of the fractal beam's 
cross-section is 
\be \label{3-cylinder}
I^{(d)}_z = \iint_A y^2 \; dA_{x}(d) ,
\ee
where $d=d_{yz}$ is the fractal dimension of 
the circular cross-section of the beam.
In Eq. (\ref{3-cylinder}) we take into account the density 
of states $c_2(d,y,z)$ in the fractal material 
through the relation $dA_{x}(d) = c_2(d,y,z) \, d A_x$,
where ($x$, $y$, $z$) are dimensionless variable. 

Let us derive the polar moment of inertia $I^{(d)}_{p}$ 
for the circular cross-section. 
By using the equalities 
\[ I^{(d)}_y=I^{(d)}_z ,\qquad I^{(d)}_p=I^{(d)}_y+I^{(d)}_z ,\]
we find the moment of inertia by using the relationship 
\be \label{yzp-1}
I^{(d)}_y=I^{(d)}_z =\frac{1}{2} \,I^{(d)}_p .
\ee

The equation for the polar moment of inertia $I^{(2)}_{p}$ 
can be written in the form
\be \label{1-0} 
I^{(2)}_p = \rho_0 \int_A (y^2+z^2) dA_{2} , \ee
where $dA_2=dydz$, ($x=x_1$, $y=x_2$, $z=x_3$) 
are dimensionless Cartesian coordinates, 
and $\rho_0$ is the constant surface mass density.

The fractional generalization of Eq. (\ref{1-0}) is 
given by expression
\be \label{1-1} 
I^{(d)}_p=\rho_0 \int_A (y^2+z^2) dA_{d} , \ee
where 
\be \label{1-2} 
dA_{d}=c(d) (\sqrt{y^2+z^2})^{d-2} dA_2 , \quad
c(d)=\frac{2^{2-d}}{\Gamma(d/2)} , \quad 0 < d \leqslant 2 . 
 \ee
Substitution of Eq. (\ref{1-2}) into Eq. (\ref{1-1}) gives
\be \label{1-6}
I^{(d)}_p= \rho_0 c(d) \int_A (y^2+z^2)^{d/2} dA_2  . 
\ee
In equation (\ref{1-1}) we use the numerical factor $c(d)$ 
such that the limits $d \to (2-0)$
give the usual integral formula (\ref{1-0}).
For $d=2$, Eq. (\ref{1-1}) gives Eq. (\ref{1-0}).
The parameter $d=d_{yz}$ denotes the fractal mass dimension 
of the circular cross-section of the beam. 
This parameter can easily be calculated from 
the experimental data by using the box counting method 
for the cross-section of the beam.

Let us now consider the circular region $A$ that is defined by 
\be \label{1-cyl}  
A=\{(y,z): \ 0\leqslant y^2+z^2 \leqslant R^2 \} . \ee
In polar coordinates $(\phi,r)$, we have
\be \label{Sr-cylinder}
dA_2 = dydz = r dr d\phi , \quad (y^2+z^2)^{d/2}=r^{d} . 
\ee
Substitution of Eq. (\ref{Sr-cylinder}) into Eq. (\ref{1-6}) gives
\be \label{1-11}
I^{(d)}_p= 2 \pi \rho_0 c(d) \, \int^R_0 r^{d+1} dr =
\frac{2 \pi \rho_0 c(d)}{(d+2)} R^{d+2} . 
\ee
This equation defines the second moment of 
the fractal beam's cross-section.
If $d=2$, we obtain the well-known equation
$I^{(2)}_p= (1/2) \pi\rho_0 R^4$.

The mass of the homogeneous fractal beam is
\be \label{1-Ma} 
M_d=\rho_0 \int_A dA_{d} , \ee
where $dA_{d}$ is defined by equation (\ref{1-2}),
and $\rho_0$ is the constant surface mass density.
Using the polar coordinates (\ref{Sr-cylinder}), 
we obtain the following mass expression  
\be \label{1-M-F}
M_d= 2 \pi \, \rho_0 \, c(d) \,\int^R_0 r^{d-1} dr =
\frac{2 \pi \, \rho_0 \, c(d)}{d} R^{d} . \ee
Substituting (\ref{1-M-F}) into (\ref{1-11}), we get 
\be \label{1-I-F} 
I^{(d)}_p=\frac{d}{d+2} M_d R^2 ,
\ee
where $d$ is the fractal mass dimension of 
the beam's circular cross-section ($1< d \leqslant 2$). 
If $d=2$, we derive the well-known relation 
$I^{(2)}_p = (1/2) M R^2$.
If we consider a fractal beam with mass and radius 
that are equal to the mass and radius of 
a beam with integer mass dimension, 
then these second moments are connected by the equation
\be 
I^{(d)}_p = \frac{2d}{d+2} \, I^{(2)}_p ,\ee
where $I^{(2)}_p$ is the moment for the homogeneous beam with 
the integer cross-section mass dimension $d=2$.

Using the relation (\ref{yzp-1}), we get
\be \label{yzp-2}
I^{(d)}=I^{(d)}_z = \frac{\pi \, \rho_0}{4}\, R^4 =
\frac{d}{2(d+2)} M_d R^2 = \frac{2d}{d+2} \, I^{(2)}_z .
\ee
This is the second moment of a circular cross-section 
of the fractal beam with cross-section 
in the $yz$-plane and fractal dimension $d=d_{yz}$, 
which should be determined by experiment.

\subsection{Gradient Euler-Bernoulli static equation for
fractal beam}

The gradient Euler-Bernoulli fractal homogeneous beam equation 
for the static case ($D^{1}_t w =0$ and $q(x,t)=q(x)$) is
obtained from Eq. (\ref{FEB-2F}) as
\be \label{FEB-ST-1}
D^{2}_x \, \Bigl( c_1(\alpha_x,x) \, D^{2}_x w \Bigr) - 
l^2_F (d) \, D^{3}_x \, \Bigl( c_1(\alpha_x,x) \, D^{3}_x w \Bigr) = 
\frac{c_1(\alpha_x,x)}{E\, I^{(d)}} \, q(x) .
\ee
For a non-fractal beam ($\alpha_x=1$),  
the static gradient Euler-Bernoulli beam equation 
takes the form
\be \label{FEB-ST-2}
D^{4}_x w - l^2_s \, D^{6}_x w  = 
\frac{1}{E\, I^{(2)}} \, q(x) .
\ee

It is noted that Eq. (\ref{FEB-ST-1}) for a fractal beam 
is analogous to the static case of Eq. (\ref{FEB-1NF}) 
for a non-fractal beam ($\alpha_x=1$
and $c_1(\alpha_x,x)=1$), which is non-homogeneous such that the product $E\, I^{(2)}_{eff}$ depends on $x$ as well as $c_1(\alpha_x,x)$, i.e. $E\, I^{(2)}_{eff} \sim x^{\alpha_x-1}$ ($0<\alpha_x<1$).  
This effective static equation for a gradient 
Euler-Bernoulli non-homogeneous beam is expressed by
\be \label{FEB-1NF2}
D^{2}_x \, \Bigl( (E\, I^{(2)}_{eff}) \, (D^{2}_x) w \Bigr) 
- l^2_s \, D^{3}_x \, \Bigl( (E\, I^{(2)}_{eff}) \,  D^{3}_x w \Bigr) = 
q_{eff}(x) 
\ee
with the effective external load 
$q_{eff}(x)= c_1(\alpha_x,x) q(x)$.

For the homogeneous case ($q(x)=0$), equation (\ref{FEB-ST-1})
can be written in the form
\be \label{FEB-ST-3}
x \, D^{4}_x w(x) + (\alpha_x-1) \, D^{4}_x w(x)
- l^{-2}_F (d) \, x \, D^{2}_x  w (x) = 
C_5 \, x^{2-\alpha} + C_6 \, x^{3-\alpha} ,
\ee
where we take into account the form of the density of states 
$c_1(\alpha_x,x)= x^{\alpha_x-1}/\Gamma (\alpha_x)$ and $x>0$.
Here $C_5$ and $C_6$ are constants defined by the 
boundary conditions for the initial problem given by
Eq. (\ref{FEB-ST-1}), which is a
differential equation of 6th order.
The general solution of Eq. (\ref{FEB-ST-3}) has the form
\[
w(x)=C_1 + C_2 \, x + C_3 \, _{1}F_{2} \Bigl[ -1/2; 1/2, \alpha_x/2-1; l^{-2}_F (d) \, x^2 /4 \Bigr] + \]
\be \label{FEB-ST-4}
+ C_4 \Bigl( l^{-1}_F (d) \, x^{2-\alpha_x/2} \, K_{\alpha_x/2-1}( l^{-1}_F (d) \, x) +
l^{\alpha-x/2-2}_F (d) \, x \, I( l^{-1}_F (d) \, x,\alpha_x)
 \Bigr) ,
\ee
where $C_1$, $C_2$, $C_3$ and $C_4$ are constants defined by 
appropriate boundary conditions; $\, _{1}F_{2} [a_1;b_1,b_2;c]$ 
denotes the hypergeometric function; 
$K_a(x)$ denotes the modified (hyperbolic) Bessel function of the second kind; and $I(x,\alpha)$ is the integral of the Bessel function of the form
\be
I(x,\alpha) = \int x^{1-\alpha_x/2} \, K_{a/2-1} (x) \, dx .
\ee
We can also use the fundamental solution for 
ordinary differential equations (2.105) 
in Kamke's book \cite{Kamke} for the case $b<0$ and $0<a<2$
where $b=- l^{\alpha-x/2-2}_F (d)$ and $a=\alpha_x-1$.

\subsection{Gradient Timoshenko equations for fractal beam}

In this section we consider a gradient generalization
of the Timoshenko beam equations for a fractal beam,
as suggested in \cite{MOS-4,MOS-5}. 
In the Timoshenko beam theory without axial effects, 
the displacement vector ${\bf u}(x,y,z,t)$ 
of the beam is assumed to be given by
\be 
u_x(x,y,z,t) = -z \, \varphi(x,t) \, \quad 
u_y(x,y,z,t) = 0 , \quad 
u_z(x,y,t) = w(x,t) ,
\ee
where $(x,y,z)$ are the coordinates of a point in the beam, 
($u_x$, $u_y$, $u_z$) are the components of the displacement vector ${\bf u}$ , 
$\varphi=\varphi(x,t)$ is the angle of rotation of the normal to the mid-surface of the beam, 
and $w=w(x,t)$ is the displacement of the mid-surface in the $z$-direction.

In \cite{MOS-4,MOS-5} it is suggested to use the derivatives 
\be \label{Der-MOS}
\partial_{x,\alpha} = c^{-1}_1(\alpha_x,x) \, D^{1}_x ,  
\quad \partial^n_{x,\alpha} =(\partial_{x,\alpha} )^n \quad 
(n \in \mathbb{N})
\ee
instead of the usual derivatives $D^1_x$ and $D^n_x$ 
for fractal materials.
If we use the derivatives given by Eq. (\ref{Der-MOS})  
for fractal materials according to \cite{MOS-4,MOS-5}, 
then the gradient Timoshenko equation for a fractal beam
can be derived from the force and moment balance equations
\be \label{FMBE-2}
\rho \, A \, D^2_t w = \partial_{x,\alpha} Q , \quad 
\rho \, I^{(d)} \, D^2_t \varphi = Q - \partial_{x,\alpha}  M ,
\ee
with the bending moment $M$ given by
\be
M = - \, E\, I^{(d)} \, \partial_{x,\alpha}  \Bigl( \varphi - l^2_s \, \partial^2_{x,\alpha} \varphi \Bigr) ,
\ee
and the shear force $Q$ is
\be
Q = k \, G \, A \, \Bigr( \partial_{x,\alpha} w - \varphi \Bigr)
- l^2_s \, k \, G \, A \, \partial^2_{x,\alpha} 
\Bigr( \partial_{x,\alpha} w - \varphi \Bigr) .
\ee

Then, the gradient Timoshenko equations for 
a homogeneous fractal beam have the form
\be \label{GTBE-1}
\rho \, A \, D^2_t w = k\, G\, A\, \partial_{x,\alpha} 
(\partial_{x,\alpha} w -\varphi )
- l^2_s \, k \, G \, A \, \partial^3_{x,\alpha}  
\Bigr( \partial_{x,\alpha} w - \varphi \Bigr) ,
\ee
\be \label{GTBE-2}
\rho \, I^{(d)} \, D^2_t \varphi = 
k \, G\, A \, (\partial_{x,\alpha} w -\varphi ) + 
E \, I^{(d)} \, \partial^2_{x,\alpha} \varphi 
- l^2_s \, k \, G \, A \, \partial^2_{x,\alpha} 
\Bigr( \partial_{x,\alpha} w - \varphi \Bigr) - E\, I^{(d)}\, l^2_s \, \partial^4_{x,\alpha}  \varphi .
\ee

The gradient Timoshenko fractal beam Eqs.
(\ref{GTBE-1}) and (\ref{GTBE-2}) can also be derived 
from an appropriate variational principle. 
The Lagrangian for a Timoshenko fractal beam with gradient non-locality has the form
\[ \mathcal{L}_{GTFB} 
= \frac{1}{2} \rho \, I^{(d)} \, \left( D^1_t \varphi(x,t) \right)^2 +
\frac{1}{2} \rho \, A \, \left( D^1_t w(x,t) \right)^2 -
\]
\[ 
- \frac{1}{2} (k G A) \, \left( \partial_{x,\alpha} w(x,t) - \varphi(x,t)  \right)^2 
 - \frac{1}{2} (E\, I^{(d)}) \, \left( \partial_{x,\alpha} \varphi(x,t) \right)^2 -
\]
\be \label{Lagr-TB-2}
- \frac{1}{2} (k G A) \, l^2_s \,
\left( \partial^2_{x,\alpha} w(x,t) - \partial_{x,\alpha} \varphi \right)^2  
- \frac{1}{2} (E\, I^{(d)}) \, l^2_s \,
\left( \, \partial^2_{x,\alpha} \varphi (x,t) \right)^2  .
\ee
Then, the stationary action principle gives 
the equations
\be \label{TB-2-5}
\frac{\partial\mathcal{L}}{\partial w} - 
D^1_t \, \left(\frac{\partial \mathcal{L}}{\partial D^1_t w}\right)  
- \, D^{1}_x \, \left(\frac{\partial \mathcal{L}}{\partial D^{1}_x w}\right)  
+ \, D^{2}_x \, \left(\frac{\partial \mathcal{L}}{\partial D^{2}_x w}\right) 
= 0 ,
\ee
\be \label{TB-2-6}
\frac{\partial\mathcal{L}}{\partial \varphi } - 
D^1_t \, \left(\frac{\partial \mathcal{L}}{\partial D^1_t \varphi }\right)  
- D^{1}_x \, \left(\frac{\partial \mathcal{L}}{\partial D^{1}_x \varphi }\right) 
+ D^{2}_x \, \left(\frac{\partial \mathcal{L}}{\partial D^{2}_x \varphi }\right) = 0 .
\ee
Equations (\ref{TB-2-5})-(\ref{TB-2-6}) 
are the Euler-Lagrange equations for the fractal beams
considered herein, as 
described by the Lagrangian given by Eq. (\ref{Lagr-TB-2}).
Substitution of Eq. (\ref{Lagr-TB-2}) into Eqs.
(\ref{TB-2-5})-(\ref{TB-2-6}) suggests that 
the gradient Timoshenko fractal beam equations 
(\ref{GTBE-1}) and (\ref{GTBE-2}) can be expressed as
\be \label{GTBE-1b}
\rho \, A \, D^2_t w = k\, G\, A\, \partial_{x,\alpha} \, \Bigl( 1 - l^2_s \, \partial^2_{x,\alpha}  \Bigr)\,
\Bigr( \partial_{x,\alpha} w - \varphi \Bigr) ,
\ee
\be \label{GTBE-2b}
\rho \, I^{(d)} \, D^2_t \varphi = 
k \, G\, A \, \Bigl(1 - l^2_s \, \partial^2_{x,\alpha} \Bigr) \,
\Bigl(\partial_{x,\alpha} w -\varphi \Bigr) + 
E \, I^{(d)} \, \partial^2_{x,\alpha} \Bigl( \varphi 
 - l^2_s \, \partial^2_{x,\alpha}  \varphi \Bigr).
\ee
If $\alpha=1$, then Eqs. (\ref{GTBE-1b})-(\ref{GTBE-2b}) 
reduce to the gradient Timoshenko equations for a beam 
made by a homogeneous non-fractal material. 

For the models based on \cite{MOS-3}-\cite{MOS-7}, 
solutions of equations for fractal materials can be obtained from solutions of equations for non-fractal materials. 
Let $w_c(x,t)$ and $\varphi_c(x,t)$  be solutions of 
Eqs. (\ref{GTBE-1b})-(\ref{GTBE-2b}) with $\alpha=1$ and $x>0$, i.e., of the gradient Timoshenko equations for homogeneous non-fractal beams. 
Then, the solutions $w_F(x,t)$ and $\varphi_F(x,t)$  
of equations (\ref{GTBE-1b})-(\ref{GTBE-2b}) for a fractal beam with $0<\alpha <1$ can be represented in terms of
$w_c$ and $\varphi_c$ as follows:
\be
w_F(x,t)= w_c(x^{\alpha}/\Gamma(\alpha+1),t), \quad
\varphi_F(x,t)=\varphi_c(x^{\alpha}/\Gamma(\alpha+1),t) .  
\ee

As an example, we consider the equation for 
an Euler-Bernoulli homogeneous fractal beam
in the absence of a transverse load ($q(x)=0$), 
\be
\rho \, A\, D^2_t w(x,t)+ E\, I^{(d)} \, \partial^4_{x,\alpha} w(x,t) = 0.
\ee 
This equation can be solved using the Fourier decomposition of 
the displacement into the sum of harmonic vibrations of the form
$w(x,t) = \text{Re}[w(x) \, \text{exp} (-i\omega t)]$,
where $\omega$ is the frequency of vibration. 
Then, for each value of frequency, we can solve 
the ordinary differential equation
\be
- \rho\, A\, \omega^2 w(x)
+E\, I^{(d)} \, \partial^4_{x,\alpha} w(x) = 0 .
\ee
The boundary conditions for a cantilevered fractal beam of length $L$ fixed at $x = 0$ are
\be \label{BC-1}
w(0)= 0 , \quad (\partial^1_{x,\alpha} w)(0) = 0 , 
\ee
\be \label{BC-2}
(\partial^2_{x,\alpha} w)(L) = 0 , \quad
(\partial^3_{x,\alpha} w)(L) = 0 .
\ee
The solution for the Euler-Bernoulli homogeneous fractal beam is defined by
\be \label{SOL-1}
w_{F,n}(x) = w_0 \Bigl(\cosh (k_n x^{\alpha}) - \cos (k_n x^{\alpha}) +
C_n(\alpha)\, [ \sin (k_n x^{\alpha}) - \sinh (k_n x^{\alpha}) ]\Bigr) , 
\quad x \in [0;L] ,
\ee
where $w_0$ is a constant, and
\be
C_n(\alpha)=\frac{\cos (k_n L^{\alpha}) + \cosh (k_n L^{\alpha})}{\sin (k_n L^{\alpha}) + \sinh (k_n L^{\alpha})} , 
\qquad
k_n = \frac{1}{\Gamma(\alpha+1)} \, \left(\frac{\rho \, A\, \omega_n^2}{E\, I^{(d)}}\right)^{1/4} .
\ee
For the boundary conditions given by 
Eqs. (\ref{BC-1})-(\ref{BC-2}), the solution (\ref{SOL-1}) 
exist only if $k_n$ are defined by
\be
\cosh(k_n L)\,\cos(k_n L) + 1 = 0 .
\ee
This trigonometric equation is solved numerically. 
The corresponding natural frequencies of vibration are
$\omega_n = k_n^2 \sqrt{(E \, I^{(d)}) /\rho \, A}$.
For a non-trivial value of the displacement, $w_0$ 
ia assumed to be arbitrary, and the magnitude of the displacement is taked as unknown for free vibrations. 
Usually, $w_0=1$ is used when plotting mode shapes.

\subsection{Combined strain-acceleration gradients for fractal beam}

Let us consider a 1D model for a fractal material that is 
described by the action 
\be \label{action-1F2}
S [w] = \int dt \, \int dl_{\alpha_x} \, 
\mathcal{L}(x, t, w, D^{1}_t w, D^{2}_x w , D^{3}_x w , D^{2}_x D^1_t w) ,
\ee
with the Lagrangian 
\[ \mathcal{L}(x, t, w, D^{1}_t w, 
D^{2}_x w , D^{3}_x w) = \frac{1}{2} \rho \, A \,  
\left( D^{1}_t w(x,t) \right)^2 +
\frac{1}{2} E\, I^{(d)} \, \left( D^{2}_x w(x,t) \right)^2 -\]
\be \label{Lag-1F2}
- \frac{1}{2} E\, I^{(d)} \, l^2_F(d)  
\left( D^{3}_x w(x,t) \right)^2  - q(x,t) w(x,t) ,
\ee
where $ dl_{\alpha_x} =  dx \, c_1(\alpha_x,x)$, 
takes into account combined 
strain-acceleration gradients \cite{AA2011}.
The stationary action principle $\delta S_F[w]=0$, 
gives the Euler-Lagrange equation in the form
\[
\frac{\partial\mathcal{L}}{\partial w} 
- c_1(\alpha_x,x) \, D^{1}_t \, \left(\frac{\partial \mathcal{L}}{\partial D^{1}_t w}\right)  
+ D^{2}_x \, \left( c_1(\alpha_x,x) \,\frac{\partial \mathcal{L}}{\partial D^{2}_x w}\right) - \]
\be \label{ELE-1F2}
- D^{3}_x \, \left( c_1(\alpha_x,x) \,\frac{\partial \mathcal{L}}{\partial D^{3}_x w}\right) 
- D^{1}_t \,D^{2}_x \, \left( c_1(\alpha_x,x) \,\frac{\partial \mathcal{L}}{\partial D^{2}_x D^1_t w}\right) = 0 .
\ee
For a homogeneous fractal beam, we obtain
\be \label{FEB-2F2}
\rho \, A \, D^{2}_t w 
+ E \, I^{(d)} \, \mathbb{D}^{4}_{x,\alpha_x}  w  
- l^2_F (d) \, E \, I^{(d)} \, \mathbb{D}^{6}_{x,\alpha_x}  w 
+ l^2_f (d) \, \rho \, I^{(d)} \, D^{2}_t \, \mathbb{D}^{4}_{x,\alpha_x}  w 
- q(x,t) = 0 ,
\ee
where the notation (\ref{not-1}) was used.

In we use the fractional continuum model 
\cite{AP2005-2,PLA2005-1,TarasovSpringer} 
with some changes suggested in \cite{MOS-3}-\cite{MOS-5},
we derive the Euler-Lagrange equation in the form 
of Eq. (\ref{FEB-2F2}), 
where the derivatives $\mathbb{D}^{2n}_{x,\alpha_x}$
are replaced by $\partial^{2n}_{x,\alpha_x} = (c^{-1}(\alpha_x,x) \, D^1_{x})^{2n}$
such that
\be \label{FEB-2F2-MOS}
\rho \, A \, D^{2}_t w 
+ E \, I^{(d)} \, \partial^{4}_{x,\alpha_x} w  
- l^2_F (d) \, E \, I^{(d)} \, \partial^{6}_{x,\alpha_x} w 
+ l^2_f (d) \, \rho \, I^{(d)} \, D^{2}_t \, 
\partial^{4}_{x,\alpha_x} w - q(x,t) = 0 .
\ee

If the beam is non-fractal, then $D=3$, $\alpha_x=1$, 
$c_1(\alpha_x,x) =1$, and Eqs. (\ref{FEB-2F2}) 
and (\ref{FEB-2F2-MOS}) take the form
\be \label{FEB-2F3}
\rho \, A \, D^{2}_t w 
+ E \, I^{(2)} \, D^{4}_x w  - 
l^2_s \, E \, I^{(2)} \, D^{6}_x  w 
+l^2_d \, \rho \, I^{(2)} \, D^{2}_t \, D^{4}_x  w 
- q(x,t) = 0 .
\ee
This is the usual combined strain-acceleration gradient 
beam model \cite{AA2011}.

Note that Eq. (\ref{FEB-2F2}) for a fractal beam 
is analogous to the equation for the usual combined 
strain-acceleration gradient non-fractal beam ($\alpha_x=1$
and $c^{-1}_1(\alpha_x,x)=1$), which is non-homogeneous, 
such that the product $E\, I^{(2)}_{eff}$ 
depends on $x$, as well as on $c_1(\alpha_x,x)$; i.e. 
$E\, I^{(2)}_{eff} \sim x^{\alpha_x-1}$ ($0<\alpha_x<1$).  
Equation (\ref{FEB-2F2-MOS}) can be solved by the method 
suggested in Section 5.5 from the solutions of 
Eq. (\ref{FEB-2F3}) for non-fractal materials.

\newpage
\section{Conclusions}

In this paper, 
we consider non-standard generalizations of 
the gradient elasticity theory \cite{A1992}-\cite{AMA2008} 
for complex materials
with power-law non-locality, long-term memory and fractality.
These non-standard generalizations may be 
important in describing unusual properties of 
nanomaterials \cite{Nano1,Nano2}.

To obtain the governing equations for the 
new fractional generalizations of gradient elasticity theory 
for materials with power-law non-locality, 
we use a new fractional variational principle 
for Lagrangians with Riesz fractional derivatives. 
New generalizations can also be obtained through 
extensions of the traditional variational calculus 
for Lagrangians by using other types of fractional 
derivatives \cite{Om-1}-\cite{Om-2}, as well as
with Riesz derivatives in the form suggested in \cite{Om-3}.
We also assume that new fractional integral elasticity models 
can be derived by using the variational principle suggested
in \cite{Int-1}, where the Lagrangian contains 
fractional integrals instead of fractional derivatives.

The fractional approach, which is suggested in this paper, 
allows us to obtain exact analytical solutions 
of the fractional differential equations 
for models of a wide class of material 
with fractional gradient non-locality.
A characteristic feature of the behavior 
of a fractional non-local continuum 
is the appearence of spatial power-tails of non-integer order.
The fractional gradient models, which are suggested 
in this paper to describe complex materials 
with fractional non-locality, can be characterized 
by a common or universal spatial behavior of elastic materials 
in analogy to the universal temporal behavior 
of low-loss dielectrics \cite{Jo1}-\cite{JPCM2008-1}.

The proposed generalization of gradient elasticity theory 
for fractal materials is based on the fractional continuum models 
proposed in \cite{PLA2005-1}-\cite{MPLB2005-1} 
(see also \cite{TarasovSpringer,MOS-1}).
In particular, equations for gradient models of fractal materials 
are obtained by a fractional integral generalization 
of the variational principle suggested in 
\cite{MPLB2005-1,Physica2005} (see also \cite{TarasovSpringer}).
In the framework of the fractional integral continuum model
for fractal materials, modified variational principles 
considered in \cite{MOS-1,MOS-2} can also be used.

We assume that new non-standard generalizations of 
the gradient elasticity models of fractal materials 
can be obtained by using the analysis on fractals 
\cite{Kugami,Strichartz-1}, as well as 
by using the methods of the vector calculus 
for non-integer-dimensional spaces \cite{CNSNS2015,JMP2014},
and by also using a generalization of
fractal lattice models \cite{JPA2008}-\cite{FL-3}.

The approach proposed in this paper is based on 
fractional integral continuum models and 
it may have a wide application
because of the relatively small numbers of parameters 
that define fractal media of great complexity and 
rich structure.
The fractional continuum model of fractal elastic materials
can be used not only to calculate global values 
and stationary characteristics, 
but also to describe dynamical 
properties of fractal materials.

\section*{Acknowledgment} 

The support of ERC-13 and the Aristeia II projects through 
the General Secretariat of Research and 
Technology (GSRT)of Greece is gratefully acknowledged.


\end{document}